\documentclass[fleqn, 11pt, letterpaper]{article}
\usepackage[english]{babel}
\usepackage{lmodern}
\usepackage{amssymb}
\usepackage{color}
\usepackage[lmargin=1in,rmargin=1in,bottom=1in,top=1in,twoside=False]{geometry}
\usepackage{esvect}
\usepackage{enumitem}
\usepackage{amsfonts}
\usepackage{amsmath}
\usepackage{amssymb}
\usepackage{amsthm}
\usepackage[english]{babel}
\usepackage{complexity}
\usepackage{mathrsfs}
\usepackage{graphics}
\usepackage[pagewise]{lineno}
\usepackage{mathtools}
\usepackage{microtype}
\usepackage[defaultlines=4,all]{nowidow}
\usepackage[many]{tcolorbox}
\usepackage[colorinlistoftodos,prependcaption,textsize=scriptsize]{todonotes}

\usepackage{xcolor}
\usepackage{xspace}
\usepackage{makecell}

\usepackage[hidelinks]{hyperref}
\usepackage[nameinlink]{cleveref}

\usetikzlibrary{calc,arrows, automata, fit, shapes, arrows.meta, positioning, decorations.pathmorphing, patterns}

\tcolorboxenvironment{theorem}{
    colframe=cyan, sharp corners=west, 
    colback=cyan!10,
	rightrule=0pt,
	toprule=0pt,
	bottomrule=0pt,
	top=0pt,
	right=0pt,
	bottom=0pt,
 }
\tcolorboxenvironment{corollary}{
    colframe=blue, sharp corners=west, 
    colback=blue!10,
	rightrule=0pt,
	toprule=0pt,
	bottomrule=0pt,
	top=0pt,
	right=0pt,
	bottom=0pt,
 }
\tcolorboxenvironment{lemma}{
    colframe=red, sharp corners=west,
	colback=red!10,
	rightrule=0pt,
	toprule=0pt,
	bottomrule=0pt,
	top=0pt,
	right=0pt,
	bottom=0pt,
 }

 \tcolorboxenvironment{result}{
    colframe=yellow, sharp corners=west, 
    colback=yellow!10,
	rightrule=0pt,
	toprule=0pt,
	bottomrule=0pt,
	top=0pt,
	right=0pt,
	bottom=0pt,
 }
\tcolorboxenvironment{proposition}{
    colframe=green, sharp corners=west, 
    colback=green!10,
	rightrule=0pt,
	toprule=0pt,
	bottomrule=0pt,
	top=0pt,
	right=0pt,
	bottom=0pt,
 }

\renewcommand{\epsilon}{\varepsilon}
\renewcommand{\phi}{\varphi}

\renewcommand{\phi}{\varphi}

\renewcommand{\epsilon}{\varepsilon}

\renewcommand{\NP}{\textsf{NP}\xspace}
\renewcommand{\FPT}{\textsf{FPT}\xspace}

\renewcommand{\PSPACE}{\textsf{PSPACE}\xspace}

\usepackage{xspace}

\newcommand{\Oof}{\mathcal{O}}

\newcommand{\Aa}{\mathcal{A}}
\newcommand{\Bb}{\mathscr{B}}
\newcommand{\Cc}{\mathscr{C}}
\newcommand{\Dd}{\mathscr{D}}

\newcommand{\Gg}{\mathscr{G}}

\newcommand{\Kk}{\mathcal{K}}

\newcommand{\Ll}{\mathcal{L}}

\newcommand{\Pp}{\mathcal{P}}

\newcommand{\Tt}{\mathscr{T}}
\newcommand{\TTT}{\mathscr{T}}

\newcommand{\Xx}{\mathcal{X}}

\newcommand{\N}{\mathbb{N}}

\newcommand{\dist}{\mathrm{dist}}

\newcommand{\strA}{\mathbb{A}}
\newcommand{\strB}{\mathbb{B}}

\renewcommand{\phi}{\varphi}

\renewcommand{\FO}{\textsf{FO}\xspace}

\newcommand{\lca}{\mathrm{lca}}

\newcommand{\MSO}{\textsf{MSO}\xspace}
\newcommand{\LFP}{\textsf{LFP}\xspace}
\newcommand{\SO}{\textsf{SO}\xspace}
\newcommand{\CMSO}{\textsf{CMSO}\xspace}

\newcommand{\NLOGSPACE}{\textsf{NLogSpace}}
\newcommand{\NEXPTIME}{\textsf{NExpTime}}
\newcommand{\PTIME}{\textsf{PTime}}

\newcommand{\conn}{\mathsf{conn}}

\newcommand{\bag}{\mathsf{bag}}
\newcommand{\cone}{\mathsf{cone}}
\newcommand{\cmp}{\mathsf{comp}}
\newcommand{\mrg}{\mathsf{mrg}}
\newcommand{\adh}{\mathsf{adh}}

\newcommand{\parent}{\mathsf{parent}}
\newcommand{\children}{\mathsf{children}}

\newcommand{\FOMSO}{\FO(\MSO(\preccurlyeq, A), \sigma)}

\newcommand{\FODP}{\ensuremath{\mathsf{FO}\hspace{0.6pt}\texttt{+}\hspace{1pt}\mathsf{dp}}}
\newcommand{\FOconn}{\textsf{FO}\hspace{0.6pt}\texttt{+}\hspace{1pt}\textsf{conn}}
\newcommand{\CMSOdp}{\textsf{CMSO/tw}\hspace{0.6pt}\texttt{+}\hspace{1pt}\textsf{dp}}

\theoremstyle{remark}
\newtheorem{theorem}{Theorem}[section]
\newtheorem{corollary}{Corollary}[section]
\newtheorem{definition}{Definition}[section]
\newtheorem{lemma}{Lemma}[section]

\crefname{corollary}{Corollary}{Corollaries}
\crefname{lemma}{Lemma}{Lemmas}
\crefname{section}{Section}{Sections}

\newtheorem*{result*}{}
\newtheorem*{remark*}{Remark}

\begin{document}

\title{Advances in Algorithmic Meta Theorems}

\date{}

\author{
Sebastian Siebertz\\
University of Bremen, Germany\\
\texttt{siebertz@uni-bremen.de}
\and 
Alexandre Vigny\\
University Clermont Auvergne, France\\
\texttt{alexandre.vigny@uca.fr}
}

\maketitle

\begin{abstract}
    \noindent Tractability results for the model checking problem of logics yield powerful algorithmic meta theorems of the form:
    \bigskip
    
    \parbox{13cm}{\emph{Every computational problem expressible in a logic $\Ll$ can be solved efficiently on every class~$\Cc$ of structures satisfying certain conditions. }}

    \bigskip
    
    \noindent The most prominent logics studied in the field are (counting) monadic second-order logic~$(\textsf{C})\MSO$, and first-order logic \FO and its extensions. 
    The complexity of \CMSO model checking in general and of \FO model checking on monotone graph classes is very well understood. 
    In recent years there has been a rapid and exciting development of new algorithmic meta theorems. 
    On the one hand there has been major progress for \FO model checking on hereditary graph classes. 
    This progress was driven by the development of a 
    combinatorial structure theory for the logically defined monadically stable and monadically dependent graph classes, as well as by the advent of the new width measure twinwidth. 
    On the other hand, new algorithmic meta theorems for new logics with expressive power between \FO and \CMSO offer a new unifying view on methods like the irrelevant vertex technique and recursive understanding. 
    In this paper we overview the recent advances in algorithmic meta theorems and provide rough sketches for the methods to prove them. 
\end{abstract}

\section{Logics Expressing Computational Problems - Descriptive \\ Complexity}

\subsection{Logics to Express Computational Problems}

Logic provides a universal and machine independent language to formally define computational problems. 
For example, the \textsc{$k$-Colorability} problem for any fixed number $k$ can be expressed by the following sentence of monadic second-order logic $\MSO$:

\vspace{-3mm}
\[\phi_k:= \exists M_1\ldots \exists M_k \big(\forall x \bigvee_{1\leq i\leq k} M_i(x) \wedge \bigwedge_{1\leq i\leq k}\forall x\forall y (E(x,y)\rightarrow \neg (M_i(x)\wedge M_i(y)))\big).\]

In this sentence, we existentially quantify $k$ sets of vertices $M_1,\ldots, M_k$ representing the~$k$ colors.
We then state that every element is contained in (at least) one of the $M_i$, that is, is assigned at least one color. 
Finally we verify that no two adjacent vertices are assigned the same color. 
We write $\strA\models \phi$ if a structure $\strA$ satisfies a sentence $\phi$, in other words, if the structure is a \emph{model} of~$\phi$. 
In this case, for a graph $G$, we have $G\models \phi_k$ if and only if $G$ is $k$-colorable. 
The algorithmic problem of testing whether a given structure is a model of a given sentence of a logic $\Ll$ is called the \emph{model checking problem} for $\Ll$. 

Second-order logic \SO can existentially and universally quantify over relations of arbitrary arity, as well as over elements of a structure, and is closed under the Boolean connectives $\wedge, \vee,\neg$. 
Monadic second-order logic \MSO is the fragment of \SO that allows quantification only over unary relations, also called monadic predicates, and
$\exists\SO$ is the existential fragment of \SO. 

Logic can express many problems, and in fact, by Fagin's celebrated theorem~\cite{fagin1974generalized,fagin1975monadic}, every $\NP$-property can be expressed by an $\exists \SO$ sentence. 
Furthermore, model checking for every fixed $\exists \SO$~sentence is in $\NP$. 
Whenever a logic $\Ll$ satisfies these two conditions for a complexity class $\Kk$: for every $\Kk$-property there is an $\Ll$ sentence defining it, and the model checking problem for every fixed sentence of $\Ll$ is in $\Kk$, we say that $\Ll$ \emph{captures} $\Kk$. 
Hence, Fagin's Theorem states that~$\exists \SO$ captures $\NP$. 
This result was extended by Stockmeyer~\cite{stockmeyer1976polynomial} who observed that full second-order logic captures the polynomial hierarchy. 
In the area of \emph{descriptive complexity theory} many other logics capturing complexity classes were studied, for example, 
transitive closure logic captures $\NLOGSPACE$ on ordered structures~\cite{Immerman87}, and least fixed-point logic $\LFP$ captures $\PTIME$ on ordered structures~\cite{immerman1982upper,vardi1982complexity}. 
Hence, the question whether $\P\neq \NP$ becomes equivalent to the purely logical question, independent of machine models, whether on ordered structures every~$\exists\SO$ formula is equivalent to an $\LFP$ formula. 
One of the main open questions in this area is whether there exists a logic $\Ll$ capturing $\PTIME$ also on unordered structures~\cite{chandra1982structure, gurevich1985logic}. 
We refer to the 
the textbook of Immerman~\cite{immerman1998descriptive} for more background.

\subsection{Data and Combined Complexity}
Note that in the capturing results above we consider the \emph{data complexity} of the model checking problem, that is, the complexity for testing the truth of a fixed formula on a given input structure. 
In the following we will be interested in the \emph{combined complexity}, where both the structure and the formula are part of the input. 
Model checking a second-order formula~$\phi$ quantifying over $x$ $d$-ary relations and $y$ elements on a structure of size $n\geq 2$ can be done in space 
$\Oof(x\cdot n^d+y\cdot \log n)\subseteq \Oof(|\phi|\cdot n^{|\phi|})$, and hence in time $2^{\Oof(|\phi|\cdot n^{|\phi|})}$, by a straight-forward algorithm that recursively iterates over all possible instantiations of the quantified relations and elements.  
Already the combined complexity of $\exists\SO$ is complete for $\NEXPTIME$. 
For~$\MSO$ space $\Oof(x\cdot n + y\cdot \log n)\subseteq \Oof(|\phi|\cdot n)$, and hence time $2^{\Oof(|\phi|\cdot n)}$ is sufficient. 
The combined complexity of $\MSO$ model checking is $\PSPACE$-complete~\cite{stockmeyer1974complexity}.
We refer to the textbooks~\cite{gradel2007finite, libkin2004elements} for more background. 

\subsection{The Complexity and Shortcomings of First-Order Logic}
First-order logic \FO uses only quantification over element variables. 
It is much weaker than the above second-order logics, but still can express many interesting properties, such as the existence of an independent set of a fixed size $k$, dominating set of fixed size $k$, and many more. 
The complexity of \FO model checking is highly relevant, as \FO forms the logical core of the database query language~SQL~\cite{codd1970relational}, and query evaluation, enumeration and counting are among the most important problems for databases. 
The existential conjunctive fragment of $\FO$ corresponds to conjunctive queries in database theory.
The data complexity of~\FO is in~$\AC^0$~\cite{barrington1990uniformity} (a circuit complexity class representing constant parallel time), while the combined complexity is \PSPACE-complete even on structures with only two elements~\cite{stockmeyer1974complexity,vardi1982complexity}. 
The combined complexity of conjunctive queries (existential conjunctive~\FO) is \NP-complete on structures with at least two elements~\cite{chandra1977optimal}.
A first-order formula with $y$ quantifiers on a structure with $n$ elements can be evaluated in space \mbox{$\Oof(y\cdot \log n)\subseteq \Oof(|\phi|\cdot \log n)$}, and hence in time $n^{\Oof(|\phi|)}$. 
Assuming the exponential time hypothesis~$\mathsf{ETH}$, this running time cannot be improved to $n^{o(|\phi|)}$~\cite{chen2006strong}.

First-order logic has two main shortcomings. 
First, it cannot express general cardinality properties (except fixed hard-coded properties up to a fixed threshold, and this shortcoming is also shared by $\MSO$). 
This has led to the extension by counting and arithmetic mechanisms, which are particularly relevant for the database community, see e.g.~\cite{benedikt2005expressive, etessami1997counting, grohe2018first, immerman1998descriptive, kuske2017first, kuske2018gaifman, schweikardt2005arithmetic, torunczyk2020aggregate} and the references therein.  
Second, \FO can express only local problems. 
For example, \FO cannot even express the algorithmically extremely simple problem of whether a graph is connected. 
This shortcoming has classically been addressed by adding transitive-closure or fixed-point operators, leading e.g.\ to the transitive closure logic \textsf{FO+TC}~(which captures \NLOGSPACE~on ordered structures~\cite{Immerman87}), and fixed-point logics (which capture \PTIME~on ordered structures~\cite{immerman1982upper,vardi1982complexity}). 

\section{A Fine-Grained View on the Model Checking Problem -- \\Parameterized Complexity}

\subsection{Parameterized Complexity and the \textsf{A}- and \textsf{W}-Hierarchy}

The distinction between combined complexity and data complexity of the model checking problems gives already some insights on the role of the formula and the structure in the complexity of the problem. 
An even finer analysis can be given when studying the \emph{parameterized complexity} of the problem. 
Parameterized complexity theory provides a multi-variate approach that takes into account additional parameters, besides the input size, that allow for a fine-grained analysis of the complexity of problems~\cite{cygan2015parameterized,DowneyF99,downey2013fundamentals, FlumG06}. 
A problem is called \emph{fixed-parameter tractable} with respect to a parameter~$k$ if it can be solved in time $f(k)\cdot n^{\Oof(1)}$, where $f$ is a computable function and $n$ is the input size. 

One natural parameter to be considered for the model checking problem is the size of the input formula $\phi$. 
However, with respect to this parameter already query evaluation for conjunctive queries in $\textsf{W}[1]$-hard and the counting problem is $\#\mathsf{W}[1]$-hard already for acyclic conjunctive queries [5]. 
The $\textsf{W}$-hierarchy is a hierarchy of parameterized complexity classes \mbox{$\FPT\subseteq \mathsf{W}[1]\subseteq \mathsf{W}[2]\subseteq \ldots \subseteq \mathsf{AW}[\star]$}, which is conjectured to be strict, and establishing hardness for one of the levels of the hierarchy is widely accepted as a proof for fixed-parameter intractability~\cite{downey1995fixed}. 
There is a second prominent hierarchy, the $\mathsf{A}$-hierarchy~\cite{flum2001fixed}, \mbox{$\FPT\subseteq \mathsf{A}[1]\subseteq \mathsf{A}[2]\subseteq \ldots \subseteq \mathsf{AW}[\star]$}. 
This hierarchy is also conjectured to be strict, and we have \mbox{$\mathsf{W}[1]=\mathsf{A}[1]$} and $\mathsf{W}[i]\subseteq \mathsf{A}[i]$ for all $i\geq 2$. 

\subsection{The Need for Structural Parameters}

The $\FO$ model checking problem yields natural complete problems for the levels of the hierarchy.  We need a bit of notation. 
By $\Sigma_0$ and $\Pi_0$ denote the set of quantifier-free formulas. For $t\geq 0$, define $\Sigma_{t+1}$ as the set of formulas 
$\exists x_1\ldots \exists x_k\, \phi$,
where \mbox{$\phi\in \Pi_t$}, and~$\Pi_{t+1}$ as the set of all formulas
$\forall x_1\ldots \forall x_k \,\phi$,
where \mbox{$\phi\in \Sigma_t$}. 
For $t,u\geq 1$, a $\Sigma_t$-formula is in~$\Sigma_{t,u}$ if all quantifier blocks after the leading existential block have length at most~$u$. 
For all $t\geq 1$, model checking for \mbox{$\Sigma_{t,1}$-formulas} is complete for~$\mathsf{W}[t]$. 
For example, the \mbox{$\Sigma_{1,1}$-formula $\exists x_1\ldots \exists x_k \big(\bigwedge_{1\leq i<j\leq k}(x_i\neq x_j \wedge \neg E(x_i,x_j))\big)$} expresses the \textsc{Independent Set} problem, the most prominent $\mathsf{W}[1]$-complete problem. 
Similarly, the $\Sigma_{2,1}$-formula $\exists x_1\ldots \exists x_k\forall y\big(\bigvee_{1\leq i\leq k} (y=x_i)  \vee \bigvee_{1\leq i\leq k} E(y,x_i)\big)$ expresses the dominating set problem, which is well-known to be $\mathsf{W}[2]$-complete. 
For all $t\geq 1$, the model checking problem for $\Sigma_t$ is complete for $\mathsf{A}[t]$. 
Note that $\Sigma_{1,1}=\Sigma_1$, which immediately implies $\mathsf{W}[1]=\mathsf{A}[1]$. 
Model checking for full $\FO$ is complete for the class $\mathsf{AW}[\star]$. 

The situation for $\MSO$ is even worse. 
Since already $3$-\textsc{Colorability}, which is expressed by a fixed \MSO-formula, is \NP-complete, \MSO model checking parameterized by formula length is $\textsf{para-NP}$-hard. 

\subsection{\MSO and the Parameters Treewidth and Cliquewidth}
Hence, the parameter $|\phi|$ alone still does not yield the desired fine-grained theory of tractability for the model checking problem. 
For a tangible theory we need to take further structural parameters into account. 
This is where graph theory enters the stage, which offers a wealth of parameters to classify the complexity of inputs. 
This classification can be lifted from graphs to general structures by considering their Gaifman graphs, or by considering the incidence encoding (see \cref{sec:gaifman}). 
In a celebrated result, Courcelle in 1990 established that every $\MSO$ definable property can be tested in linear time on graphs of bounded treewidth~\cite{courcelle1990monadic}. 
More precisely, he established that the model checking problem for $\MSO$ is fixed-parameter tractable with respect to the parameters formula length and treewidth, that is, solvable in time $f(|\phi|, tw)\cdot n$, where $tw$ is the treewidth of the input graph and~$f$ is a computable function.
This result extends to counting MSO, $\CMSO$~\cite{courcelle1990monadic}, and to graphs with bounded cliquewidth~\cite{courcelle2000linear}, that is, $\CMSO$ model checking can be solved in time $f(|\phi|, cw)\cdot n$, where $cw$ is the cliquewidth of the input graph and $f$ is a computable function.
Furthermore, it was proved that efficient $(\mathsf{C})\MSO$ model checking cannot be extended to classes of unbounded treewidth or cliquewidth, assuming further mild closure conditions and the standard complexity theoretic assumptions~\cite{courcelle2007vertex,ganian2014lower,kreutzer2012parameterized,kreutzer2010lower,kreutzer2010brambles,makowsky2003tree}. 

\subsection{More Parameters for \FO}
In 1996 Seese~\cite{seese1996linear} established that FO model checking is linear time solvable on graphs with bounded maximum degree, that is, solvable in time $f(\Delta, |\phi|)\cdot n$, and thereby initiated the fine-grained study of the parameterized complexity of first-order model checking with respect to structural parameters. 
His result was extended to more and more general classes of sparse graphs~\cite{flum2001fixed,frick2001deciding,dawar2007locally,dvovrak2010deciding, kreutzer2011algorithmic,GroheKS17}. 
The last result of Grohe, Kreutzer and Siebertz~\cite{GroheKS17} shows that every FO definable property of graphs is decidable in nearly linear time on every nowhere dense class of graphs. 
Nowhere dense classes of graphs are very general classes of sparse graphs~\cite{nevsetvril2011nowhere} and turn out to be a tractability barrier for FO model checking on monotone classes (that is, classes that are closed under taking subgraphs): if a monotone class of graphs is not nowhere dense, then testing first-order properties for inputs from this class is as hard as for general graphs~\cite{dvovrak2010deciding, kreutzer2011algorithmic}. 

\subsection{Algorithmic Meta Theorems}
Note that the nature of these results is different from those in descriptive complexity theory. 
They have a much more algorithmic, rather than a complexity theoretic flavor. 
In particular, the development for the model checking problems parallels that of the development in parameterized complexity, which in the early days was strongly driven by the development of the graph minors theory of Robertson and Seymour~\cite{robertson1983graph}, and the rise of treewidth as one of the most important structural parameters. 
The model checking results not only stand by themselves but additionally capture the essence and limits of fundamental algorithmic techniques, such as dynamic programming and compositionality, the locality based method, and game based methods. Newer results are strongly based on recursive understanding and the irrelevant vertex technique. 
In their seminal surveys Martin Grohe and Stephan Kreutzer coined the expression \emph{algorithmic meta theorems}~\cite{grohe2008logic,grohe2009methods,kreutzer2011algorithmic}. 

\bigskip
    
    \parbox{13cm}{\emph{Every computational problem expressible in a logic $\Ll$ can be solved efficiently on every class~$\Cc$ of structures satisfying certain conditions. }}

    \bigskip

Since the groundbreaking work of Courcelle~\cite{courcelle1990monadic}, algorithmic meta theorems have become an essential tool in the toolbox of parameterized complexity theory. 
Formalizing a problem in a logic yields a fast and convenient, and yet formal proof for its tractability on certain classes of structures. 
At this time, one of the main goals in the area is to find the most general classes of structures that allow for fixed-parameter tractable \FO model checking. 
A recent trend is also to find new logics with expressive power between that of \FO and \MSO that capture certain algorithmic techniques. 

\subsection{The Limits of Tractability}

As already mentioned, the complexity of \CMSO model checking is essentially settled. 
For plain~\MSO the question is related to a conjecture of Seese~\cite{seese1991structure}, stating that if a class of graphs has a decidable satisfiability problem for \MSO, then it has bounded cliquewidth. 
A stronger version conjectures (formulated in the contrapositive) that if a class $\Cc$ has unbounded cliquewidth, then \MSO can encode all graphs in $\Cc$, which essentially yields intractable \MSO model checking on classes of unbounded cliquewidth. 
This statement for \CMSO is implied by a connection between \CMSO and vertex minors~\cite{courcelle2007vertex}. 

The tractability limit of \FO model checking on monotone graph classes is constituted by nowhere dense graph classes. 
Consequently, attention has shifted to study more general, hereditary classes (that is, classes that are closed under taking induced subgraphs).

\subsection{New Directions for \FO Model Checking}

Recently, there has been very exciting progress on algorithmic meta theorems, which essentially goes into three directions. 
The first direction is inspired by the classical \emph{interpretation method} (see \cref{sec:interpretation-method}). 
The interpretation method allows to lift results from a class $\Cc$ to classes $\Dd$ that can be logically interpreted in $\Cc$. 
The classical proof of Courcelle's Theorem can be seen as a prime example of the interpretation method: every graph $G$ with a fixed bounded treewidth (or with a fixed bounded cliquewidth) can be encoded in a colored tree $T^+$ such that \MSO can recover $G$ from~$T^+$. 
The $(\mathsf{C})$\MSO model checking problem for classes of bounded treewidth (or cliquewidth) then reduces to model checking on colored trees, which is very easy to solve. 
When it is possible to encode a class $\Dd$ in a sparse class~$\Cc$ via an \FO transduction, we call $\Dd$ a \emph{structurally sparse class}. 
Note that it is still a challenge to recover a sparse preimage from a dense but structurally sparse input graph. 
Results for bounded degree graphs were lifted to classes with structurally bounded degree~\cite{gajarsky2020new,gajarsky2018recovering}, from bounded expansion classes to classes with structurally bounded expansion~\cite{gajarsky2020first} and from nowhere dense classes to structurally nowhere dense classes~\cite{dreier2023first}. 

The second direction was inspired by a result of Adler and Adler~\cite{adler2014interpreting}, who observed that nowhere dense classes are \emph{monadically dependent (NIP)} and \emph{monadically stable}. 
The notions of dependence and stability are key notions from classical model theory~\cite{shelah1990classification} and are defined by forbidden interpretable configurations. 
This connection to model theory brought new notions of structural tameness and a large toolbox for infinite theories to graph theory. 
The challenge to develop a combinatorial and algorithmic theory for finite graphs was quickly accepted and led to many nice results, see e.g.~\cite{bonnet2022model,braunfeld2022decomposition,dreier2021lacon,dreier2023first2,dreier2022treelike,DreierMMSV22,dreier2023first,dreier2023indiscernibles,dreier2024flip,gajarsky2020first,GajarskyMMOPPSS23,gajarsky2022stable,jiang2020regular,mahlmann2024monadically,malliaris2014regularity,nevsetvril2021rankwidth,nevsetvril2020linear,ohlmann2023canonical,torunczyk2023flip}.

A class is monadically dependent if in colorings of the graphs from the class one cannot interpret all graphs; a natural candidate for the limit of tractability for \FO model checking, and soon it was conjectured that monadic dependence constitutes the tractability boundary for \FO model checking on hereditary classes ~\cite{bonnet2024twin,dreier2023first,gajarsky2020new,warwick16}. 
A big step towards this conjecture was taken by Dreier et al.~\cite{dreier2023first2,dreier2023first} who proved that model checking is tractable on monadically stable classes, which are important subclasses of monadically dependent classes.
Furthermore, the hardness part of the conjecture was established in~\cite{dreier2024flip}.

Some indication for the truth of the conjecture is given by the third direction. 
Bonnet et al.~\cite{bonnet2021twin} introduced the new notion of \emph{twinwidth}, which had an immense impact in structural graph theory in the past few years. 
Bonnet et al.\ proved that \FO model checking is tractable on classes with bounded twinwidth assuming that we are given a contraction sequence, witnessing the boundedness of twinwidth, together with the input. 
It turns out that a hereditary class of ordered graphs has bounded twinwidth if and only if it is monadically dependent~\cite{bonnet2024twin}. 
Furthermore, on ordered structures bounded twinwidth contraction sequences can be efficiently computed, and model checking on hereditary ordered classes beyond bounded twinwidth is intractable~\cite{bonnet2024twin}. 
Hence, the conjecture is true on ordered graphs
The main challenge in the area remains to resolve the conjecture for general hereditary classes. 

\smallskip
One consequence of the model checking result on classes with bounded twinwidth that is seldomly mentioned is the tractability of \emph{order-invariant} \FO on classes of bounded twinwidth (assuming that contraction sequences are given with the input). 
An order-invariant formula may use a symbol for a linear order and must satisfy the semantic property that its truth on ordered structures is independent of the chose order. 
Note that it is undecidable to decide if a formula is order-invariant~\cite{libkin2004elements}, order-invariance must be guaranteed for the formulas that are input to the model checking problem.  
Order-invariant \MSO is more expressive than \CMSO~\cite{GanzowR08} but the model checking problem is tractable on the same classes as for \CMSO~\cite{eickmeyer2020model}. 
An unpublished result of Gurevich states that
the expressive power of order-invariant \FO is stronger than that of plain \FO
(see, e.g., Theorem~5.3 of~\cite{libkin2004elements} for a presentation of the result).
Model checking for the weaker successor-invariant \FO is tractable on classes with bounded expansion~\cite{eickmeyer2020model}, for order-invariant \FO the limit of tractability is wide open. 

\subsection{Why do we not study the other classical logics}

The fact that first-order logic can only express local problems is classically addressed by adding transitive-closure or fixed-point operators~\cite{eflum,gradel2007finite,Libkin04}.  
These logics however, do not have a rich algorithmic theory: 
Even the model checking problem for the very restricted monadic transitive-closure logic
$\mathsf{TC}^1$ is \textsf{AW}$[\star]$-hard on planar graphs of maximum degree at most $3$~\cite{grohe2008logic}. 
Model checking for full $\SO$ is intractable even on colored paths, as \PSPACE-computations on strings can be simulated.

\subsection{\FO plus connectivity, compound logic, and more}

This has motivated recent approaches to introduce new logics whose expressive power lies between~\FO and \MSO, which are tailored to express algorithmic graph problems and which are still tractable on interesting graph classes. 
Examples include \emph{Separator logic}~\cite{bojanczyk2021separator,schirrmacher2023first}, \emph{Disjoint paths logic}~\cite{schirrmacher2023first}, \emph{Compound logics for modification problems}~\cite{fomin2023compound}, and \CMSO/\textsf{tw} as the fragment of \MSO on graphs obtained by appropriately restricting set quantification~\cite{sau2024parameterizing}.
These logics can express many problems that are studied in parameterized complexity, and hence provide very useful meta theorems. 
Again, these logics are not only relevant for the meta theorems they provide, but they also capture important algorithmic paradigms. 

\subsection{Advances in Algorithmic Meta Theorems}

It has been about 15 years since Grohe and Kreutzer wrote their seminal expositions on algorithmic meta theorems~\cite{grohe2008logic,grohe2009methods,kreutzer2011algorithmic}. 
They identified the following key methods. 
\vspace{1mm}
\begin{enumerate}
    \item The Automata Theoretic Method, translating \CMSO formulas into automata that can be run on trees. 
    \item The Reduction or Interpretation Method, allowing to translate tractability results between classes of structures via logical interpretations. 
    \item The Composition Method, which allow to lift results to structures that are composed of simpler pieces from these pieces.
    \item Locality Based Methods for \FO, exploiting that \FO can express only local properties. 
    \item Coloring and Quantifier-Elimination, which allows to simplify to quantifier-free formulas by an algorithmic enrichment of the signature. 
\end{enumerate}

Of course, these foundational methods have not changed over time. In this paper we aim to give an overview over new techniques and the recent extensions of the well-established methods. 

The automata method has been generalized from trees to 
trees that are additionally augmented with graph structures~\cite{PilipczukSSTV22}. 
This allows in many cases to directly combine the composition method with the automata method. 

The composition method has gained new momentum with the advent of the method of recursive understanding developed in parameterized complexity~\cite{chitnis2016designing,kawarabayashi2011minimum}, and its predecessor, the decomposition into unbreakable parts~\cite{cygan2020randomized,cygan2014minimum}. 
The technique has led to the result of Lokshtanov et al.~\cite{LokshtanovR0Z18}, showing that the problem of deciding a \CMSO property on general graphs can be reduced to the same problem on unbreakable graphs. 
It is also the basis for the model checking result for separator logic~\cite{PilipczukSSTV22} and disjoint paths logic~\cite{schirrmacher2024model} on classes with excluded topological minors, as well as for model checking of compound logic~\cite{fomin2023compound} and $\CMSO/\mathsf{tw}$~\cite{sau2024parameterizing} on classes with excluded minors. 

The locality based methods have been taken to their limits by combining locality with recursive game based decompositions of local neighborhoods on nowhere dense classes~\cite{GroheKS17} and monadically stable classes~\cite{dreier2023first,dreier2023first2}. 
An appropriate extension to monadically dependent classes is one of the main open problems in the area. 

The advent of twinwidth has led to a completely new foundational method of dynamic programming along contraction sequences, see e.g.~\cite{bonnet2022twinwin,bonnet2021twin2,BonnetG0TW21,bonnet2024twin,BonnetGMT23,DBLP:conf/soda/BonnetKRT22,bonnet2022twin,bonnet2021twin,geniet2024twin}. 


\subsection{Structure of this paper}

We take this rapid and exciting development as a motivation to give a brief overview over the new methods for algorithmic meta theorems. 
We structure the paper as follows. 
After fixing our notation we first introduce the interpretation method and quantifier elimination as special cases of parameterized reductions between model checking problems. 
These remaining base cases are the extended automata based method, locality based methods, and methods based on twinwidth. 



\section{Preliminaries}

\subsection{Structures}

Let us recall some basics from model theory. We refer to the textbooks~\cite{Ebbinghaus94,gradel2007finite,Hodges93,libkin2004elements} for extensive background. 
A \emph{signature} is a collection
of relation and function symbols, each with an associated arity. Let $\sigma$ be a
signature. A {\em $\sigma$-structure} $\strA$ consists of a non-empty
set $A$, the \emph{universe} of~$\strA$, together with an interpretation of
each $k$-ary relation symbol $R\in\sigma$ as a $k$-ary relation
$R_\strA\subseteq A^k$ and an interpretation of each $k$-ary function symbol $f\in \sigma$ as a $k$-ary function $f_\strA:A^k\rightarrow A$. 
In this work we assume that all structures are finite (i.e.\ have a finite universe and a finite signature).
We write $|\strA|$ for the size of the universe of $\strA$ and $\|\strA\|$ for the size of an encoding of $\strA$, e.g.\ in the standard incidence encoding. 

\medskip
In this work we mostly consider the following types of structures:
\begin{itemize}
    \item \emph{Colored graphs} are $\sigma$-structures, where $\sigma$ consists of a single binary relation symbol $E$ and unary relation symbols, with the property that
	$E$ is interpreted as a symmetric and anti-reflexive relation. 
    \item \emph{Guided pointer structures} are $\sigma$-structures, where $\sigma$ consists of a single binary relation symbol~$E$, unary relation symbols, and unary function symbols, with the property that $E$ is interpreted as a symmetric and anti-reflexive relation, and that the interpretation of every function $f\in\sigma$ is \emph{guided}, meaning that for a graph $G$ if $f_G(u)=v$, then $u=v$ or $uv\in E(G)$. 
    \item \emph{Ordered graphs} are $\sigma$-structures, where $\sigma$ consists of two binary relation symbols $E$ and $<$, with the property that $E$ is interpreted as a symmetric and anti-reflexive relation, and $<$ is interpreted as a linear order. 
    \item \emph{Partial orders} are $\sigma$-structures, where $\sigma$ consists of a binary relation symbol $\sqsubseteq$ that is interpreted as a partial order, in many cases a linear order or tree order. 
\end{itemize}

\subsection{First-order and monadic second-order logic}

We first define first-order logic \FO and monadic second-order logic \MSO (over the signature~$\sigma$). 
The fancier logics will be defined in the sections where they are first discussed. 
We assume an infinite supply $\textsc{Var}_1$ of first-order variables and an infinite supply $\textsc{Var}_2$ of monadic second-order variables. 
Every variable is a term, and if $t_1,\ldots, t_k$ are terms and $f\in \sigma$ is a $k$-ary function symbol, then also $f(t_1,\ldots, t_k)$ is a term. 
$\FO[\sigma]$ formulas are built from the atomic formulas $t_1=t_2$, where~$t_1$ and~$t_2$ are terms, and $R(t_1,\ldots, t_k)$, where \mbox{$R\in \sigma$}
is a $k$-ary relation symbol
and $t_1,\ldots, t_k$ are terms, by closing under the Boolean
connec\-tives~$\neg$,~$\wedge$~and~$\vee$, and by existential and
universal quantification~$\exists x$ and $\forall x$. 

Monadic second-order formulas are defined as first-order formulas, but further
allow the use of monadic quantifiers $\exists X$ and $\forall X$, and of a membership atomic formula $x\in X$, where~$x$ is a first-order variable and $X$ a monadic second-order variable. 
\CMSO is the extension of \MSO with modulo counting quantifiers $\exists^{a[b]}x\phi(x)$, expressing the existence of $a$ modulo $b$ elements satisfying~$\phi$.

A variable~$x$ not in the scope of a quantifier is a {\em free variable} (we do not consider formulas with free set second-order variables). A formula without free variables is a {\em sentence}.
The {\em quantifier rank} $\mathrm{qr}(\varphi)$ of a formula $\varphi$ is the
maximum nesting depth of quantifiers in~$\phi$. 
A formula without quantifiers is called {\em quantifier-free}.

If $\strA$ is a $\sigma$-structure
with universe $A$, then an {\em assignment} of variables in~$\strA$
is a mapping $\bar a:\textsc{Var}_1 \rightarrow A$. We use the standard
notation $(\strA, \bar a)\models \phi(\bar x)$ or $\strA \models \phi(\bar a)$
to indicate that $\phi$ is satisfied in $\strA$ when the free variables $\bar x$
of $\phi$ have been assigned by $\bar a$. 
We write~$\models\phi$ to express that~$\phi$ is a valid sentence, that is, $\phi$ holds in every structure (of an appropriate signature). 
For a formula $\phi(\bar x)$ we define $\phi(\strA):=\{\bar a\in A^{|\bar x|} \mid \strA\models\phi(\bar a)\}$.

Two structures $\strA$ and $\strB$ are \emph{$(\Ll,q)$-equivalent}, written $\strA\equiv_q^\Ll \strB$, if they satisfy the same $\Ll$-sentences of quantifier rank at most $q$. 
The \emph{$(\Ll,q)$-type} of an element $a$ in $\strA$ is the collection of all formulas~$\phi(x)$ of quantifier rank at most $q$ such that $\strA\models\phi(a)$.
A set of formulas is a \emph{$q$-type} if it is the $q$-type of some element in some structure.

\subsection{Interpretations and transductions}

Let $\sigma,\tau$ be relational signatures. 
An \emph{$\Ll$-interpretation} $\mathsf{I}$ of $\sigma$-structures in $\tau$-structures is a tuple $\mathsf I=(\nu(\bar x), (\rho_R(\bar x_1,\ldots, \bar x_{ar(R)}))_{R\in \sigma})$, where $\nu(\bar x)$ and $\rho_R(\bar x_1,\ldots, \bar x_{ar(R)})$ are $\Ll$-formulas, and 
$|\bar x|=|\bar x_i|$ for all $1\leq i\leq ar(R)$ for all $R\in \sigma$, where $ar(R)$ is the arity of $R$.  
For every $\tau$-structure~$\strB$, the $\sigma$-structure $\strA=\mathsf I(\strB)$
has the universe~$\nu(\strB)$ and each relation~$R_\strA$ is interpreted as $\rho_R(\strB)$. 
We say that $\nu$ \emph{defines} the vertex set and $\rho_R$ defines the relation~$R$ of $\mathsf{I}(\strB)$. 
The number of free variables $|\bar x|$ of $\nu$ is the \emph{dimension} of the interpretation. 
A $1$-dimensional interpretation is also called a \emph{simple interpretation}. 


An \emph{$\Ll$-transduction} consists of the replacement of structure $\strB$ by the union of a fixed number of disjoint copies of $\strB$, augmented with relations between the copies, followed by a monadic lift, which consists of marking some subsets of vertices with new unary predicates, and finally a simple $\Ll$-interpretation. 
Let us formalize these notions. 

Let $k\in\mathbb N$. The \emph{$k$-blowing} of a $\tau$-structure $\strB$ with universe $B$ is the $\tau'$-structure $\strB \bullet k$,
where~$\tau'$ is the signature obtained from $\tau$ by adding a new binary relation $\sim$ encoding an equivalence relation. 
The domain of $\strB\bullet k$ is $B\times[k]$, and, denoting $\pi$
the projection $B\times[k]\rightarrow B$ we have, for all $x,y\in B\times [k]$,
$\strB\bullet k\models x\sim y \Leftrightarrow \pi(x)=\pi(y)$, and (for $R\in\tau$)
$\strB\bullet k\models R(x_1,\dots,x_k) \Leftrightarrow \strB\models R(\pi(x_1),\dots,\pi(x_k))$.

A \emph{monadic lift} of a $\tau$-structure $\strB$ is a $\tau^+$-expansion
$\strB^+$ of $\strB$, where $\tau^+$ is the union of~$\tau$ and a
set of unary relation symbols. 

For $k\in \N$, an \emph{$(\Ll,k)$-transduction} of $\sigma$-structures from $\tau$-structures is a pair $\mathsf T=(\mathsf I, k)$, where~$\mathsf I$ is a simple $\Ll$-interpretation of $\sigma$-structures in $\tau^+$ structures, where $\tau^+$ is an extension of $\tau'$ (the extension of $\tau$ by the symbol $\sim$) by unary relation symbols.

Let $\mathsf T=(\mathsf I,k)$ be an $(\Ll,k)$-transduction. A $\sigma$-structure $\strA$ can be $\mathsf T$-transduced from a $\tau$-structure~$\strB$ if there is a $\tau^+$-lift $\strB^+$ of $\strB\bullet k$ such that $\strA=\mathsf I(\strB^+)$. 
A class $\Cc$ of $\sigma$-structures can be $\mathsf T$-transduced from a class $\Dd$ of $\tau$-structures if for every structure $\strA\in \Cc$ there exists a structure~$\strB\in \Dd$ such that~$\strA$ can be $\mathsf T$-transduced from~$\strB$. 
A class $\Cc$ of $\sigma$-structures can be transduced from a class~$\Dd$ of $\tau$-structures if it can be
$\mathsf T$-transduced from $\Dd$ for some $(\Ll,k)$-transduction~$\mathsf T$.
In case~$k=1$ we speak of a copyless transduction


\subsection{Gaifman graphs and incidence encodings}\label{sec:gaifman}

The \emph{Gaifman graph} of a $\sigma$-structure $\strA$ with universe $A$ is the graph with vertex set $A$, where~$u$ and~$v$ are adjacent if they belong
jointly to some relation $R_\strA$ for $R\in \sigma$, or if one is the image of the other by some function $f_\strA$ for $f\in \Sigma$. 

For a relational signature $\sigma$, the \emph{incidence graph} of a $\sigma$-structure $\strA$ with universe $A$ is the colored graph with vertices $A$ and one vertex for each tuple $\bar v$ appearing in a relation~$R_\strA$ (we take multiple copies if a tuple appears in several relations). 
If $\bar v=(v_1,\ldots, v_k)$ and $\bar v\in R_\strA$, then the vertex $\bar v$ in the incidence graph is marked with a color $R$ and connected with an edge of color $i$ with the vertex~$v_i$ for $1\leq i\leq k$. 
Note that $\strA$ is interpretable by a simple ninterpretation from its incidence graph, but vice versa, in general we need a higher-dimensional interpretation to interpret the incidence graph of a structure from the structure. 

In uniformly sparse structures the structure is transduction equivalent to its Gaifman graph and to its incidence graph. 
More precisely, when the \emph{star chromatic number} of the Gaifman graph of a structure is bounded, then the structure is bi-transducible with its Gaifman graph via copyless transductions~\cite[Lemma 3.1]{bonnet2024twinperm}, and the structure is bi-transducible with its incidence graph. 
The second statement is not explicit in~\cite{bonnet2024twinperm}, however, it follows by an easy modification of the proof of Lemma 3.1 of that paper. 
We refer to~\cite{bonnet2024twinperm} for the definition of the star chromatic number and remark that all classes with bounded treewdith, and even all classes with bounded expansion have bounded star chromatic number. 
Note that transductions compose, and, as every \FO-transduction is also an \MSO-transduction, both \FO- and \MSO-transductions over structures whose Gaifman graphs have bounded star chromatic number have the same expressive power as over their incidence encodings. 
As a consequence of this fact and the interpretation method (\cref{sec:interpretation-method}) we get for example that Courcelle's Theorem for treewidth also holds for guarded \MSO (in particular $\MSO_2$), the extension of \MSO with quantification over subsets of relations. 

Hence, we can define the structural complexity of a sparse class of structures as the structural complexity of the class of its Gaifman graphs. 
For dense classes we have to relate to the class of its incidence graphs.

\section{Reductions Between Algorithmic Meta Theorems}

Recall that the model checking problem for $\Ll$ on a class $\Cc$ of structures is the problem: given $\strA\in \Cc$ and $\phi\in\Ll$, decide whether $\strA\models\phi$. We denote it by MC$(\Ll,\Cc)$. 

\begin{definition}
A \emph{non-uniform algorithmic meta theorem} is a result of the form: Let $\Ll$ be a logic and~$\Cc$ a class of structures. Then for all sentences $\phi\in \Ll$ there exists an algorithm that given $\strA\in \Cc$ decides whether $\strA\models\phi$ in time $f(\phi)\cdot \|\strA\|^{\Oof(1)}$ for some function $f$. 
\end{definition}

\begin{definition}
An \emph{algorithmic meta theorem} is a result of the form: Let $\Ll$ be a logic and~$\Cc$ be a class of structures. Then \emph{MC}$(\Ll,\Cc)$ is fixed-parameter tractable, that is, there is an algorithm that on input $\strA\in \Cc$ and $\phi\in \Ll$ decides whether $\strA\models\phi$ in time $f(\phi)\cdot \|A\|^{\Oof(1)}$ for a computable function~$f$. 
\end{definition}

We define reductions between algorithmic meta theorems simply as parameterized reductions between the corresponding model checking problems. 

\begin{definition}
    Let \emph{MC}$(\Ll_1,\Cc)$ and \emph{MC}$(\Ll_2,\Dd)$ be model checking problems. 
    A \emph{parameterized reduction} from \emph{MC}$(\Ll_1,\Cc)$ to \emph{MC}$(\Ll_2,\Dd)$ is a function $R$ that maps instances $(\strA,\phi)$ of \emph{MC}$(\Ll_1,\Cc)$ to instances $(\strB,\psi)$ of \emph{MC}$(\Ll_2,\Dd)$ such that 
    \vspace{1mm}
    \begin{enumerate}
        \item $\strA\models\phi\Leftrightarrow \strB\models\psi$, 
        \item $|\psi|\leq f(|\phi|)$ for a computable function $f$, and 
        \item $(\strB,\psi)$ can be computed in time $g(|\phi|)\cdot \|(\strA,\phi)\|^{\Oof(1)}$ for a computable function $g$. 
    \end{enumerate}
    
    We say that \emph{MC}$(\Ll_1,\Cc)$ reduces to \emph{MC}$(\Ll_2,\Dd)$, and write \emph{MC}$(\Ll_1,\Cc) \leq_{fpt}$\, \emph{MC}$(\Ll_2,\Dd)$, if there exists a parameterized reduction from \emph{MC}$(\Ll_1,\Cc)$ to \emph{MC}$(\Ll_2,\Dd)$.
\end{definition}

We now have the standard reduction lemma. 

\begin{lemma}
    Let \emph{MC}$(\Ll,\Cc)$ and \emph{MC}$(\Ll',\Cc')$ be model checking problems with \emph{MC}$(\Ll,\Cc) \leq_{fpt}$\, \emph{MC}$(\Ll',\Cc')$. 
    If \emph{MC}$(\Ll',\Cc')$ is fixed-parameter tractable, then \emph{MC}$(\Ll,\Cc)$ is fixed-parameter tractable. 
\end{lemma}

\subsection{Example: Reducing Separator Logic to \FO with \MSO Atoms on Augmented Trees}

As already discussed above, \FO falls short of being able to express algorithmic problems that involve \emph{non-local} properties. 
For example, $\FO$ cannot express the very simple algorithmic question whether two vertices are connected. 
\emph{Separator logic}, denoted by $\FOconn$ and independently introduced in~\cite{bojanczyk2021separator, schirrmacher2023first}, enriches $\FO$ with connectivity predicates that are tailored to express algorithmic graph properties that are commonly studied in parameterized algorithmics. 
Separator logic is obtained from \FO by adding 
the atomic predicates $\conn_k(x,y,z_1,\ldots, z_k)$ that hold true in a graph if there exists a path between (the valuations of) $x$ and $y$ after (the valuations of) $z_1,\ldots, z_k$ have been deleted. Separator logic   can express many interesting problems such as the \textsc{Feedback Vertex Set} problem and \textsc{Elimination Distance} problems to first-order definable classes.
It was proved in \cite{PilipczukSSTV22} that model checking for separator logic is fixed-parameter tractable on classes excluding a topological minor, and for subgraph-closed classes, this result cannot be extended to more general classes (assuming a further condition on the efficiency of encoding required for the hardness reduction). 

Separator logic yields a first nice example of the reduction method. 
First, it was observed in~\cite{PilipczukSSTV22} that separator logic on highly connected graphs can be reduced to plain \FO. 
Let us formalize the concept of high connectivity. 

A \emph{separation} in a graph $G$ is a pair $(A,B)$ of vertex subsets such that $A\cup B=V(G)$ and there are no edges in $G$ between $A\setminus B$ and $B\setminus A$. 
The {\em{order}} of the separation is the cardinality of the {\em{separator}}~$A\cap B$. 
A vertex subset $X$ is {\em{$(q,k)$-unbreakable}} in a graph $G$ if for every separation $(A,B)$ in~$G$ of order at most $k$ in~$G$, either $|A\cap X|\leq q$ or $|B\cap X|\leq q$. 
Intuitively, a separation of order~$k$ cannot break $X$ in a balanced way: one of the sides must contain at most $q$ vertices of $X$.

Now observe that for a $(q,k)$-unbreakable graph $G$, the query $\conn_k(u,v,x_1,\ldots,x_k)$ can be expressed in plain \FO. 
The query fails if and only if there is a set $A$ of at most $q$ vertices that contains exactly one of the vertices $u$ and $v$, and such that all neighbors of vertices of~$A$ outside of~$A$ are contained in $\{x_1,\ldots,x_k\}$. 
The existence of such a set of $q$ vertices can be expressed using $q$ existential quantifiers followed by a universal quantifier.
So every formula that uses only $\conn_k$ predicates can be rewritten as a plain $\FO$ formula on $(q,k)$-unbreakable graphs, as long as $q$ is a constant. Denote by $\FOconn_k$ the fragment of $\FOconn$ that uses only $\conn_k$-predicates and denote by $\Bb_{q,k}$ the class of $(q,k)$-unbreakable graphs.

\begin{lemma}\label{lem:foconn-fo}
    MC\,$(\FOconn_k,\Bb_{q,k}) \leq_{fpt}$\, MC\,$(\FO,\Bb_{q,k})$.
\end{lemma}

Now, for general graphs, we reduce separator logic to \FO with \MSO atoms, denoted \mbox{$\FO(\MSO(\preccurlyeq))$}, over augmented trees, using the following decomposition theorem. 
Let us first define trees and tree decompositions. 

A (rooted) tree is an acyclic and connected graph $T$ with a distinguished root vertex~$r$. 
We write $\parent(x)$ for the parent of a node~$x$ of $T$, and $\children(x)$ is the set of children of~$x$
in~$T$. 
We define $\parent(r)=\bot$. 
A vertex $x\in V(T)$ is an ancestor of a vertex $y\in V(T)$, written $x\preceq_T y$, or simply~$x\preceq y$ if $T$ is clear from the context, if $x$ lies on the unique path between $y$ and the root $r$.
Note that hence every node is an ancestor of itself.  
For nodes $x,y\in V(T)$, we write $\lca(x,y)$ for the least common ancestor of $x$ and $y$ in $T$. 
Note that $x$ is an ancestor of $y$ if and only if $\lca(x,y)=x$. 

A {\em{tree decomposition}} of a graph $G$ is a pair $\mathcal{T}=(T,\bag)$, where $T$ is a rooted tree and
$\bag\colon V(T)\to 2^{V(G)}$ is a mapping that assigns to each node
$x$ of $T$ a {\em{bag}} $\bag(x)\subseteq V(G)$, such that 
\vspace{1mm}
\begin{itemize}
\item for every $u\in V(G)$, the set of nodes $x\in V(T)$ satisfying $u\in \bag(x)$ induces a connected and nonempty subtree of $T$, and 
\item for every edge $uv\in E(G)$, there exists a node $x\in V(T)$
  such that $\{u,v\}\subseteq \bag(x)$.
\end{itemize}

The \emph{treewidth} of a graph $G$ is the size of a largest bag of $\mathcal{T}$ minus $1$, where $\mathcal{T}$ ranges over all tree decompositions of $G$. 

\medskip
Let $\mathcal T=(T,\bag)$ be a tree decomposition and let $x\in V(T)$. 

\begin{itemize}
\item The {\em{adhesion}} of $x$ is
 $$\adh(x)\coloneqq \bag(\parent(x))\cap \bag(x).$$
 \item The {\em{margin}} of $x$ is
 $$\mrg(x)\coloneqq \bag(x)\setminus \adh(x).$$
\item The {\em{cone at $x$}} is
 $$\cone(x)\coloneqq \bigcup_{y\succeq_T x} \bag(y).$$
\item The {\em{component at $x$}} is
 $$\cmp(x)\coloneqq \cone(x)\setminus \adh(x)= \bigcup_{y\succeq_T x} \mrg(y).$$
\end{itemize}


Observe that the margins $\{\mrg(x)\colon x\in V(T)\}$ are pairwise disjoint and cover the whole vertex set of $G$. 
The {\em{adhesion}} of a tree decomposition $\mathcal T=(T,\bag)$ is defined
as the largest size of an adhesion, that is,
$\max_{x\in V(T)}|\adh(x)|$.

A tree decomposition $\mathcal T=(T,\bag)$ of a graph $G$ is
{\em{strongly $(q,k)$-unbreakable}} if for every $x\in V(T)$,   $\bag(x)$ is $(q,k)$-unbreakable in $G[\cone(x)]$.

\begin{theorem}[\cite{cygan2019minimum}]\label{thm:strong-unbreakability}
  There is a function $q(k)\in 2^{\Oof(k)}$ such that for every graph
  $G$ and number $k$ there exists a strongly $(q(k),k)$-unbreakable
  tree decomposition of $G$ of adhesion at most $q(k)$.  Moreover,
  given~$G$ and~$k$, such a tree decomposition can be computed in time
  $2^{\Oof(k^2)}\cdot |G|^2\cdot \|G\|$.
\end{theorem}

The theorem opens two directions to approach the model checking problem for separator logic. 
The first approach is to use compositionality (see \cref{sec:compositionality}) and do dynamic programming over unbreakable tree decompositions. 
The second approach, which we follow here, works as follows. 
We encode the decomposition in a colored tree with additional edges between siblings (the children of some node) in the tree. 
We call such a structure an \emph{augmented tree}.
We will then reduce model checking of separator logic to model checking on a logic over augmented trees.
The resulting problem can then be solved via the automata method (see \cref{sec:automata}). 

More generally we may want to encode general $\sigma$-structures. In this case we represent augmented trees as relational structures equipped with the ancestor relation~$\preceq_T$ of a tree~$T$, as well as relations~$R$ relating the children of any given node, with which the tree is augmented.
We will access the global connectivity of the tree with \MSO that speaks only about~$\preceq$, as well as over the colors of the tree. We write $\MSO(\preceq, A)$ for the set of $\MSO$ formulas over such signatures, where $A$ is a finite set of colors (unary predicates). 
As we will translate such formulas into tree automata, we will often call~$A$ an alphabet. 
We now consider {\FO} formulas with restricted {\MSO} atoms. For the signature $\sigma$ and the alphabet $A$, the logic $\FO(\MSO(\preccurlyeq, A), \sigma)$
denotes first-order logic over signature~$\sigma$ where one can use formulas of $\MSO(\preceq,A)$ as atomic formulas. 
Denote by $\Gg$ the class of all graphs and by~$\mathscr{T}_{A,\sigma}$ the class of $(A,\sigma)$-augmented trees. 

\begin{theorem}[\cite{PilipczukSSTV22}]\label{thm:foconn-augmented}
    MC\,$(\FOconn,\Gg) \leq_{fpt}$\, MC\,$(\FO(\MSO(\preccurlyeq, A), \sigma),\mathscr{T}_{A,\sigma})$, where $A$ is a finite alphabet and $\sigma$ is a signature of colored graphs. 
\end{theorem}

\subsection{Example: Reducing \CMSO/\textsf{tw}\texttt{+}\textsf{dp} on Classes with Excluded Minors to \CMSO on Classes with Bounded Treewidth}

One of the classic and important problems that separator logic cannot express~\cite{schirrmacher2023first} is the \textsc{Disjoint Paths} problem: 
\textsl{Given a graph $G$ and a set $\{(s_{1},t_{1}),\ldots,(s_{k},t_{k})\}$ of pairs of terminals, the question is whether $G$ contains vertex-disjoint paths joining $s_{i}$ and~$t_{i}$ for $1\leq i\leq k$.} 
A straight-forward approach to deal with this shortcoming is to add a predicate expressing exactly this property. 
\emph{Disjoint-paths logic} $\FODP$ is an extension of separator logic that was introduced in~\cite{schirrmacher2023first}. 
It extends first-order logic~(\FO) with atomic predicates $\DP_k[(x_1, y_1), \ldots ,(x_k, y_k)]$ expressing the existence of internally vertex-disjoint paths between $x_i$ and $y_i$, for  $1\leq i\leq k$. 
Disjoint paths logic can express many interesting algorithmic problems, such as the disjoint paths problem, minor containment, topological minor containment, $\mathcal{F}$-topological minor deletion, and many more (see~\cite{golovach2023model}). 
It was shown in~\cite{golovach2023model} that model checking for disjoint-paths logic is fixed-parameter tractable on classes with excluded minors and in 
\cite{schirrmacher2024model} that it is fixed-parameter tractable on classes with excluded topological minors.

Another logic recently introduced logic by Sau, Stamoulis and Thilikos~\cite{sau2024parameterizing} is \CMSO/\textsf{tw}, a fragment of counting monadic second-order logic that allows only restricted quantification of sets in \CMSO formulas. 
Recall that \CMSO with no restriction on set quantification is intractable beyond graphs of bounded cliquewidth (under the standard complexity theoretic assumptions and mild closure conditions). 
\CMSO/\textsf{tw} very elegantly generalizes another 
extension of \FO that was introduced by Fomin et al.~\cite{fomin2023compound}, the so-called \emph{compound logic}, which is tailored to express general families of graph modification problems.
\CMSO/\textsf{tw} is very expressive, nevertheless, it cannot express the disjoint paths problem. 
Following the approach of disjoint paths logic, we may simple add an operator to express this property to obtain the logic \CMSOdp. 

Let $G$ be a graph and $X\subseteq V(G)$. 
A graph $H$ is an \emph{$X$-rooted minor} of $G$ if there is a collection $\mathcal{B} = \{B_x \mid x\in V(H)\}$ of pairwise disjoint connected subsets of $V(G)$, each containing at least one vertex of $X$, 
and such that, for every edge $xy\in E(H)$, there are~$u\in B_x$ and $v\in B_y$ with $uv\in E(G)$. 
The set $B_x$ is called the \emph{branch set} of $x$ in $G$. 
A graph~$H$ is a minor of $G$ if it is a $V(G)$-rooted minor of $G$. 
Given a graph $G$ and $X\subseteq V(G)$, the \emph{annotated treewidth} of $X$ in $G$, denoted $\mathsf{tw}(G, X)$, is the maximum
treewidth of an $X$-rooted minor of $G$~\cite{thilikos2023excluding}. 

\CMSO/\textsf{tw} is the restriction of \CMSO where instead of using the quantifier $\exists X$ (resp.\ $\forall X$) for a set variable $X$, we have quantifiers $\exists_k X$ (resp.\ $\forall_k X$) for some number $k$, where $\exists_k X$ and $\forall_k X$ mean that the quantification is applied on vertex or edge sets $X$ with $\mathsf{tw}(G,X)\leq k$. 
For the case of quantification on an edge set $X\subseteq E(G)$, $\mathsf{tw}(G,X)\leq k$ means that the set of the endpoints of the edges in $X$ has annotated treewidth at most $k$.
\CMSOdp~is the extension of \CMSO/\textsf{tw} with the disjoint paths operators.

\begin{theorem}[\cite{sau2024parameterizing}]\label{thm:cmsodp}
    Let $\Cc$ be a class excluding some minor. 
    Then there exists a class $\Dd$ of structures with bounded treewidth such that 
    MC\,$(\CMSOdp,\Cc) \leq_{fpt}$\, MC\,$(\CMSO, \Dd)$. 
\end{theorem}

The theorem is proved using the \emph{irrelevant vertex technique}, which was first introduced in~\cite{robertson1986graph}. 
In particular, it uses the flat wall theorem, in the recent refined formulation of~\cite{sau2024more}. 
A similar scheme was also applied in the works~\cite{fomin2023compound,fomin2023algorithmic,golovach2023model}. 
We stress that the irrelevant vertex technique in the proof of \cref{thm:cmsodp} does not simply remove vertices from the input graph. 
In particular, the theorem does not yield a graph (structure) of bounded treewidth that is logically equivalent to the input graph. 
This would yield a contradiction with the undecidability of the satisfiability problem of \CMSO (and \FO). 
It is crucial in the reduction that also the input formula is rewritten, depending on the input graph. 

\medskip
The model checking algorithm for \FODP~on classes with excluded topological minors combines the approaches of \cite{PilipczukSSTV22} and \cite{sau2024parameterizing}. 
First, the input graph is decomposed into unbreakable parts using \Cref{thm:strong-unbreakability}. 
On each part, we distinguish two cases. 
When a part excludes a minor, we can apply the result of~\cite{sau2024parameterizing} and iteratively remove irrelevant vertices until we arrive at a graph with bounded treewidth. 
When a part contains large minors, we can prove a generalization of \cref{lem:foconn-fo} for disjoint paths logic.

\begin{lemma}
    $\mathrm{MC}$\,$(\FODP,\Cc) \leq_{fpt}$\, MC\,$(\FO,\Cc)$, where $\Cc$ is any class of graphs that is $(q,k)$-unbreakable and contains large clique-minors, where $k,q$ and ``large'' depends on the input formula. 
\end{lemma}

The lemma is based on the combinatorics of a variant of the ``\textsl{generic folio lemma}'' proved by Robertson and Seymour in~\cite{robertson1995graph}, which was used by Grohe et al.~\cite{grohe2011finding} in order to show that testing topological minor containment is fixed-parameter tractable.
Model checking for disjoint paths logic then proceeds by dynamic programming over the tree decomposition into unbreakable parts, using to compositionality method (\cref{sec:compositionality}) to combine the solutions of the unbreakable parts into a global solution.


\section{The Interpretation Method}\label{sec:interpretation-method}

The interpretation method is a special case of the reduction method that is fundamental for many model checking algorithms. 
It is based on encoding structures from a class $\Cc$ in structures from a class $\Dd$ such that the structures from $\Cc$ can be recovered from $\Dd$ by a logical interpretation. 
This makes the translation of the formula $\phi$ in the reduction particularly elegant and simple. 
This translation is based on the following interpretation lemma. 

\begin{lemma}\label{lem:interpretation-lemma}
    Let $\mathsf I$ be an $\Ll$-interpretation of $\sigma$-structures in $\tau$-structures. 
    Then for every formula $\phi\in \Ll[\sigma]$ there exists a formula $\psi\in \Ll[\tau]$ with the following property. Let $\strA$ be a $\sigma$-structure and $\strB$ be a $\tau$-structure such that $\strA\cong\mathsf I(\strB)$. Then 
    $\strA\models \phi \Leftrightarrow \strB\models \psi$. Furthermore, $\psi$ can be efficiently computed from $\psi$. 
\end{lemma}

In fact, it is very easy to compute the formula $\psi$ from $\phi$. If $\mathsf I=(\nu, (\rho_R)_{R\in \sigma})$, we simply have to replace all occurrences of atoms $R(\bar x_1,\ldots, \bar x_{ar(R)})$ by their defining formulas $\rho_R(\bar x_1,\ldots, \bar x_{ar(R)})$, and restrict quantification to those tuples that satisfy the formula $\nu$. 


We can now define reductions that are based on interpretations. 

\begin{definition}
    A class $\Cc$ of $\sigma$-structures admits \emph{efficient} $\Ll$-encoding in a class $\Dd$ of \mbox{$\tau$-structures} if there exists an $\Ll$-interpretation $\mathsf I$ of $\sigma$-structures in $\tau$-structures and an algorithm that given $\strA\in \Cc$ and $\phi\in \Ll$ computes a structure $\strB\in \Dd$ with $\mathsf I(\strB)\cong \strA$ in time$f(|\phi|)\cdot \|\strA\|^{\Oof(1)}$ for some computable function $f$.     
\end{definition}

The next lemma is immediate by \cref{lem:interpretation-lemma} and is the key to the interpretation method. 

\begin{lemma}\label{lem:interpretation-method}
    If a class $\Cc$ of $\sigma$-structures admits \emph{efficient} $\Ll$-encoding in a class $\Dd$ of \mbox{$\tau$-structures}, then \emph{MC}$(\Ll,\Cc) \leq_{fpt}$\, \emph{MC}$(\Ll,\Dd)$. 
    Hence, if $\emph{MC}(\Ll,\Dd)$ is fixed-parameter tractable, then so is \emph{MC}$(\Ll,\Cc)$. 
%
\end{lemma}

\medskip
The interpretation method combines well with other methods. 
In many cases the signature of~$\Dd$ is chosen such that the logic we are interested in collapses to a simpler logic, for example, the signature may allow for quantifier elimination, such that model checking becomes trivial. 
Also the single steps in the game based approaches on nowhere dense and monadically stable classes can be seen as simplifications of the model checking problem via the interpretation method. 

\subsection{Example: Efficient \CMSO-Encoding on Classes with Bounded Cliquewidth in Colored Trees}

We recall the classical reduction of \CMSO on classes with bounded cliquewidth to \CMSO on colored trees. 
We define clique expressions with respect to more general base classes, as the methods for $\FO(\MSO(\preccurlyeq, A), \sigma)$ on augmented trees extends to these more general graph classes. 
We remark however, that while cliquewidth and clique decompositions can be efficiently computed~\cite{oum2006approximating}, this is not clear for clique decompositions over more general classes and we may have to require such decompositions to be given with the input. 

\smallskip
A $k$-colored graph is a graph with each vertex assigned a color from $[k]$. 
On $k$-colored graphs we define the following operations with respect to a base class $\Cc$.
\vspace{1mm}
\begin{itemize}
    \item \emph{Create} a $k$-colored graph $G$ from $\Cc$. 
    \item For a function $c:[k]\rightarrow [k]$ \emph{recolor} the vertices of the input graph according to $c$. 
    \item For a set $S$ of $2$-element subsets of $[k]$, \emph{join} a family of $k$-colored graphs by taking their disjoint union and for each $\{i,j\}\in S$ (possibly $i=j$), add an edge between every pair of vertices that have colors $i$ and $j$, respectively, and originate from different input graphs.
\end{itemize}

A \emph{width-$k$} clique decomposition is a (rooted) tree $T$ where the nodes are labeled by the above operation names in an arity preserving way,
that is, all constants are leaves and all recolor operations have 
exactly one child. 
Note that the join operation is a commutative operation, so the tree does not need to have an order on siblings. 
A clique decomposition defines the k-colored graph obtained by 
evaluating the operations in the decomposition. 

The \emph{cliquewidth} of a graph is the minimum number $k$ for which there is a width-$k$ clique decomposition whose result is (some coloring of) the graph over $\Cc_1$, which contains only the single-vertex graph. 

We remark that this definition of cliquewidth is different from the original definition~\cite{courcelle2000upper}, however, it is within factor $2$ of that definition. 
Also note that every class of graphs with bounded treewidth also has bounded cliquewidth. 

\begin{lemma}\label{lem:mso-cw}
Let $\Cc$ be a class with bounded cliquewidth. Then there exists a class $\Tt$ of colored trees such that $\Cc$ can be \MSO-encoded in $\Tt$. In particular,  
    \emph{MC}$(\CMSO,\Cc) \leq_{fpt}$\, \emph{MC}$(\CMSO,\Tt)$. 
\end{lemma}

The idea is to use the tree $T$ from the clique decomposition as the host graph. The vertices of~$G$ are the leaves of $T$. 
In order to decide whether two vertices $u,v$ are connected by an edge we have to find the color of the vertices at the least common ancestor $x=\lca(u,v)$ and check whether vertices of these colors are connected by the join operation at $x$. 
\MSO can keep track of how colors change through the tree interpret the connections accordingly. 

\subsection{Example: Structurally sparse graph classes}

With the good understanding of sparse graph classes it is a natural question to extend these results to classes that interpret or transduce in sparse classes. 
For example, the concepts of treedepth and treewidth as well as their dense analogs shrubdepth and cliquewidth can be defined in terms of transductions. 
We write $\TTT_d$ for the class of trees of depth at most $d$ and~$\TTT$ for the class of all trees. 

\begin{itemize}
    \item A class $\Cc$ of graphs has bounded treedepth if and only if the class of incidence graphs of graphs from $\Cc$ can be FO- or MSO-transduced from $\TTT_d$ for some $d\geq 1$ (this follows from the work of Ganian et al.~\cite{ganian2019shrub,ganian2012trees}).\\[-6mm]
    \item A class $\Cc$ of graphs has bounded treewidth if and only if the class of incidence graphs of graphs from $\Cc$ can be MSO-transduced from $\TTT$ by a classical result of Courcelle~\cite{courcelle1992monadic}. \\[-6mm]
    \item A class $\Cc$ of graphs has bounded treewidth if and only if the class of incidence graphs of graphs from $\Cc$ can be FO-transduced from the class of all (finite) tree orders as shown by Colcombet~\cite{colcombet2007combinatorial}. 
\end{itemize}

The dense analogs are obtained as follows. 

\begin{itemize}
    \item A class $\Cc$ of graphs has bounded shrubdepth if and only if $\Cc$ can be FO- or MSO-transduced from $\TTT_d$ for some $d\geq 1$ as shown by Ganian et al.~\cite{ganian2019shrub,ganian2012trees}.\\[-6mm]
    \item A class $\Cc$ of graphs has bounded cliquewidth if and only if $\Cc$ can be MSO-transduced from~$\TTT$ as proved by Courcelle~\cite{courcelle1992monadic}. \\[-6mm]
    \item A class $\Cc$ of graphs has bounded cliquewidth if and only if $\Cc$ can be FO-transduced from the class of all (finite) tree orders as shown by Colcombet~\cite{colcombet2007combinatorial}. 
\end{itemize}

When applied to the classical notions of sparsity we obtain the following notions of \emph{structural sparsity}, see~\cite{gajarsky2020first,nevsetvril2016structural}

\begin{itemize}
    \item A class $\Dd$ has \emph{structurally bounded degree} if there exists a class $\Cc$ of bounded degree such that $\Dd$ can be transduced from $\Cc$~\cite{gajarsky2020new,gajarsky2018recovering}. \\[-6mm]
    \item A class $\Dd$ has \emph{structurally bounded expansion} if there exists a class $\Cc$ of bounded expansion such that $\Dd$ can be transduced from $\Cc$~\cite{gajarsky2020first}. \\[-6mm]
    \item A class $\Dd$ is \emph{structurally nowhere dense} if there exists a nowhere dense class $\Cc$ such that~$\Dd$ can be transduced from $\Cc$~\cite{dreier2023first,gajarsky2020first}. 
\end{itemize}

Structurally sparse classes have a rich combinatorial theory, however, the difficulty to obtain algorithmic meta theorems for these classes lies in the problem to find the sparse pre-images of graphs when given only the dense input graph from $\Dd$. 
Only for classes with structurally bounded degree we have a meta theorem whose proof is based on the interpretation method of \cref{lem:interpretation-method}. 

\begin{theorem}[\cite{gajarsky2020new}]
    Let $\Dd$ be a class with structurally bounded degree. Then there exists a class~$\Cc$ with bounded degree such that $\Dd$ can be \FO-encoded in $\Cc$. In particular, \emph{MC}$(\FO,\Dd) \leq_{fpt}$\, \emph{MC}$(\FO,\Cc)$. 
\end{theorem}

Even though we know how to solve the \FO model checking even on monadically stable classes~\cite{dreier2023first2}, efficient sparsification of classes with structurally bounded expansion and structurally nowhere dense classes (both of these are monadically stable) remains an important open problem. 

\section{Quantifier elimination}

 Quantifier elimination is a classical technique from model theory to demonstrate the tameness of theories. 
 In the algorithmic context of finite structures we can first enrich the signature, and then eliminate quantifiers, making quantifier elimination a special case of a reduction. 
 Model checking of quantifier-free formulas is then trivial, as we just have to check the atomic type of an input tuple. 

 \subsection{Example: Quantifier Elimination on Classes with Bounded Expansion}

 We demonstrate the principle with the by now classical example of quantifier elimination on classes with bounded expansion.
 The first building block is quantifier elimination on classes with bounded treedepth. 
 To generalize to classes with bounded expansion we need a slightly stronger statement than just the quantifier elimination result. 
 Recall from the preliminaries that a guided pointer structures is a $\sigma$-structures, where $\sigma$ consists of a single binary relation~$E$, unary relations, and unary functions, with the property that $E$ is symmetric  and anti-reflexive, and that every function $f\in\sigma$ is \emph{guided}, meaning that if $f(u)=v$, then $u=v$ or $uv\in E(G)$. 

 \begin{theorem}[\cite{dvovrak2013testing}]\label{thm:qe-trees}
     For every FO-formula $\varphi(\bar x)$ and every class $\Cc$ of colored graphs with bounded treedepth there exists a quantifier-free formula $\tilde\phi(\bar x)$ and a 
     linear time computable map $Y$ such that, for every $G\in\Cc$, $Y(G)$ is a guided expansion of $G$ such that for all tuples of vertices $\bar v$
 	we have $
 	G\models \varphi(\bar v)\Leftrightarrow Y(G)\models\tilde\varphi(\bar v)$.
 \end{theorem}

 This quantifier elimination result lifts to classes with bounded expansion by so-called low treedepth colorings.
 For a positive integer $p$, a \emph{$p$-treedepth coloring} of a graph $G$ is a vertex coloring of~$G$ such that the subgraph induced by any $i\leq p$ color classes has treedepth at most $i$. 

 \begin{lemma}[\cite{zhu2009colouring}]
     A class $\Cc$ of graphs has bounded expansion if and only if for every~$p$ there exists $c(p)$ such that every $G\in \Cc$ admits a $p$-treedepth coloring with at most $c(p)$ colors. 
     Furthermore, given $G$ and $p$, such a coloring is efficiently computable. 
\end{lemma}

Using low treedepth decompositions, one can now iteratively eliminate quantifiers in bounded expansion classes. 
Note that because the computed expansions are guided by the original graph structure, the class (of Gaifman graphs) does not loose the property of having bounded expansion in each quantifier elimination step. 
This leads to the following theorem. 

\begin{theorem}[\cite{dvovrak2013testing,gajarsky2020first,grohe2009methods}]\label{thm:qe-BE}
    Let $\Cc$ be a class with bounded expansion. 
    There exists a class~$\Dd$ of guided pointer structures such that for every $G\in \Cc$ and \FO-formula $\varphi(\bar x)$ one can efficiently compute a quantifier-free formula $\tilde\phi(\bar x)$ and a 
    guided expansion $Y(G)$ of $G$ such that for all tuples of vertices $\bar v$
	\[
	G\models \varphi(\bar v)\quad\Leftrightarrow\quad Y(G)\models\tilde\varphi(\bar v).
	\]
    Consequently, \emph{MC}$(\FO,\Cc) \leq_{fpt}$\, \emph{MC}$(\mathsf{QF}, \Dd)$, and in particular, 
    as \emph{MC}$(\mathsf{QF}, \Dd)$ is fixed-parameter tractable, \emph{MC}$(\FO,\Cc)$ is fixed-parameter tractable. 
\end{theorem}

Nowhere dense classes can be characterized by low treedepth colorings similarly as bounded expansion classes. 
A class $\Cc$ of graphs is nowhere dense if and only if for every~$p$ and every $\epsilon>0$ there exists $c(p,\epsilon)$ such that every $n$-vertex graph $H\subseteq G\in \Cc$ admits a $p$-treedepth coloring with at most $c(p,\epsilon)\cdot n^\epsilon$ colors.
Note that these are too many colors to also obtain quantifier elimination on nowhere dense classes. 
In fact, it was proved that quantifier elimination as for bounded expansion classes cannot exist for nowhere dense classes~\cite{GroblerJMSV24}. 

Classes with structurally bounded expansion are characterized by the existence of $p$-shrubdepth colorings with $c(p)$ colors for each fixed $p$~\cite{gajarsky2020first} (here we do not require that the shrubdepth is at most $p$ (note that this is not even defined) but only that it is bounded for every $p$). 
Monadically stable classes are characterized by the existence of $p$-shrubdepth colorings with $c(p,\epsilon)\cdot n^\epsilon$ colors for each $p$~\cite{braunfeld2024decomposition}. 

\section{The composition method}\label{sec:compositionality}

The composition method is a very classical method for \FO and \MSO, so we only give a very rough sketch and refer to the literature~\cite{grohe2009methods,grohe2008logic,kreutzer2011algorithmic}.  
The method allows to deduce the truth of formulas in structures that are composed of simpler parts, e.g.\ via tree decompositions with small adhesion. 
The most classical composition theorems goes back to Feferman and Vaught~\cite{feferman1967first}. 
This key lemma is easily proved using Ehrenfeucht-Fra\"iss\'e games (in particular it extends to \CMSO). 

\begin{theorem}[Feferman and Vaught~\cite{feferman1967first}]
    Let $\strA,\strB$ be $\sigma$-structures and let $\bar c$ be constant symbols naming all elements of $V(\strA)\cap V(\strB)$. 
    Then the $q$-type (for \FO or \MSO) of $\strA\cup\strB$ is determined by the $q$-type of $(\strA,\bar c)$ and the $q$-type of $(\strB,\bar c)$. 
\end{theorem}

By dynamic programming along tree decompositions one obtains for example efficient \CMSO model checking on classes with bounded treewidth, Courcelle's famous theorem. 

\begin{theorem}[Courcelle's Theorem \cite{courcelle1990graph}]
    Let $\Cc$ be a class of bounded treewidth. 
    Then \emph{MC}$(\CMSO,\Cc)$ is fixed-parameter tractable. 
\end{theorem}

We obtain as an immediate corollary of \Cref{thm:cmsodp} that \emph{MC}$(\CMSOdp,\Cc)$ is fixed-parameter tractable on classes with excluded minors. 

\begin{corollary}[\cite{sau2024parameterizing}]
    Let $\Cc$ be a class excluding some minor. 
    Then 
    \emph{MC}$(\CMSOdp,\Cc)$ is fixed-parameter tractable. 
\end{corollary}

Another recent application of the composition method is the reduction of \CMSO model checking to unbreakable graphs by the recursive understanding technique~\cite{LokshtanovR0Z18}. 

\begin{theorem}[\cite{LokshtanovR0Z18}]
    Let $\phi$ be a \CMSO sentence. For all $k$ there exists $q$ such that if there exists an algorithm that solves the model checking problem for $\phi$ on $(q,k)$-unbreakable graphs in time $\Oof(n^d)$ for some $d>4$, then there exists an algorithm that solves the model checking problem for $\phi$ on general graphs in time $\Oof(n^d)$. 
\end{theorem}

Note that the result is for individual \CMSO properties. 
Efficient model checking for a single sentence $\phi$ on unbreakable graphs is a much weaker requirement than the requirement that \CMSO types (up to some quantifier rank) have to be efficiently computable on unbreakable graphs. 
The theorem is proved by recursively replacing large unbreakable parts of the graph that are glued by small separators by small parts that have the same type just with respect to $\phi$. 
This is possible for each fixed formula $\phi$, as representatives for equivalence classes can be hardcoded in the algorithm. As a consequence, the theorem only yields non-uniform fpt algorithms. 

Uniform algorithms via dynamic programming exist for many special cases, in particular model checking for disjoint paths logic on classes with excluded topological minors is fixed-parameter tractable and this result is proved using the composition method. 

\begin{theorem}[\cite{schirrmacher2024model}]
 Let $\Cc$ be a class with excluded topological minors. Then \emph{MC}$(\FODP,\Cc)$ is fixed-parameter tractable. 
\end{theorem}

\section{The automata based method}\label{sec:automata}

Also the automata theoretic method is a classical method for model checking. 
Every \CMSO property of colored trees can be translated into an equivalent automaton, which can simply be run on the tree to evaluate~\cite{doner1970tree,thatcher1968generalized}. 
As a consequence, we immediately get the following theorem. 

\begin{theorem}[\cite{doner1970tree,thatcher1968generalized}]
    Let $\Tt$ be a class of colored trees. 
    Then \emph{MC}$(\CMSO,\Tt)$ is fixed-parameter tractable. 
\end{theorem}

In order to conclude the fixed parameter tractability of model checking for separator logic on classes with excluded topological minors via the reduction of \cref{thm:foconn-augmented}, we now consider automata for $\FO(\MSO(\preccurlyeq, A), \sigma)$ that run on augmented trees. 
Recall that in augmented trees we have a tree~$T$ represented by the ancestor relation $\preccurlyeq$ and additionally labeled with letters from an alphabet $A$, as well as $\sigma$-structures on the children of inner nodes of $T$. 
In the following we will call the structure below a node $x$ the \emph{whorl} of $x$. 
The whorls are used to represent the bags of tree decompositions, for which we want to do first-order model checking (there is a small caveat here that we will comment on below). 
We define automata that read augmented trees in the standard bottom-up fashion and whose transition function is defined by first-order formulas. 
At any node $x$, to determine the state of the automaton at $x$, we consider the graph induced on the children of $x$ in the augmented tree, vertex-labeled by the states already determined in the bottom-up computation. 
Then the state at $x$ is determined by evaluating a fixed collection of first-order sentences on this labeled graph. 
Finally, the automaton accepts the augmented tree depending on the state at the root.
As a result we get that $\FO(\MSO(\preccurlyeq, A), \sigma)$ model checking is fixed parameter tractable on trees that are augmented with structures on which \FO model checking is fixed-parameter tractable. 
\pagebreak

Formally, an \emph{automaton $\Aa$ on $(A,\sigma)$-augmented trees} consists of:
\begin{itemize}
\item an input \emph{alphabet} $A$: the automaton will process augmented trees over the alphabet $A$,
\item a finite set of \emph{states} $Q$,
\item a set of \emph{accepting states} $F\subset Q$,
\item for each state $q\in Q$ and letter $a\in A$, a first-order
  sentence $\delta_{q,a}$ in the signature $\Sigma\cup Q$, where each
  $q\in Q$ is viewed as a unary relation symbol, such that for every
  fixed $a\in A$, the sentences $(\delta_{q,a})_{q\in Q}$ are mutually
  inconsistent and their disjunction $\bigvee_{q\in Q}\delta_{q,a}$ is
  equivalent to true. The sentences $\delta_{q,a}$ are called the
  \emph{transition sentences} of $\Aa$.
\end{itemize}

A \emph{run} of $\Aa$ on an augmented tree $T$ is a labeling
$\rho: V(T)\rightarrow Q$ such that for every node~$v$ with letter~$a$ and
whorl $\strA_v$, the state $q=\rho(v)$ is the unique state such that
$\delta_{q,a}$ holds in the \mbox{$Q$-labeled} structure~$\strA_v$
with labeling $\rho$ restricted to $\strA_v$.  Formally, $\strA_v$
is viewed as a structure over the signature~\mbox{$\Sigma\cup Q$},
where a predicate $q\in Q$ holds on a vertex $v$ if and only if
$\rho(v)=q$.  The run is \emph{accepting} if it labels the root with
an accepting state. The automaton~$\Aa$ \emph{accepts} an augmented
tree~$T$ if it has an accepting run on it.

\begin{theorem}[\cite{PilipczukSSTV22}]\label{thm:formulas to automata}
  For every formula $\phi(\bar x)$ of $\FOMSO$ there is an automaton~$\Aa$ on augmented trees that is equivalent to $\phi$. 
  Moreover,
  $\Aa$ is computable from $\phi$.
\end{theorem}

We call the model checking problem for classes of structures that are additionally labeled with unary predicates the \emph{labeled model checking problem}. 

\begin{lemma}[\cite{PilipczukSSTV22}]
  Let $\Cc$ be a class of\, $\sigma$-structures such that labeled model checking for \FO is
  fixed-parameter tractable on $\Cc$.
   Then automata evaluation on $\Cc$-augmented
  trees is efficient.
\end{lemma}

As a corollary we get the following theorem. 
\begin{theorem}[\cite{PilipczukSSTV22}]\label{thm:formula evaluation}
  Let $\Cc$ be a class of\, $\sigma$-structures such that labeled model checking for \FO is
  fixed-parameter tractable on $\Cc$.
    Then model-checking for \mbox{$\FOMSO$} is
  efficient on $\Cc$-augmented trees with labels from a finite alphabet $A$.
\end{theorem}

As a final corollary we obtain the following theorem. 

\begin{theorem}[\cite{schirrmacher2023first}]
    Let $\Cc$ be a class excluded a topological minor. 
    Then \emph{MC}$(\FOconn,\Cc)$ is fixed-parameter tractable. 
\end{theorem}

Let us comment on why we do not obtain fixed-parameter tractability on nowhere dense classes. 
The reason is that we were slightly imprecise when we stated that used the parts of the decomposition into unbreakable parts as the whorls in the augmentations. 
In fact, we need to toke the so-called \emph{bag graphs}, which are obtained from the subgraphs induced by a bag by turning all adhesions into cliques. 
If the original graph excludes some topological minor, then the bag graphs also exclude some (larger) topological minor, hence, \FO model checking is tractable on the bag graphs. 
If however, the original graph comes from a nowhere dense class, the bag graphs are possibly no longer nowhere dense. 

\section{Locality and Game-based Decompositions of Local Neighborhoods}

A key property of \FO that is often exploited for efficient model checking is its \emph{locality}. 
This is formalized for example by Gaifman's Locality Theorem~\cite{gaifman1982local}, which states that every
first-order formula~$\varphi(\bar x)$ is equivalent to a Boolean combination of 
\vspace{1mm}
\begin{enumerate}
\item \emph{local formulas} $\psi^{(r)}(\bar x)$ and 
\item \emph{basic local 
formulas} $\exists x_1\ldots \exists x_k\big(\bigwedge_{i\neq j}
\dist(x_i,x_j)>2r \wedge \chi^{(r)}(x_i)\big)$. 
\end{enumerate} 

Here, the notation $\psi^{(r)}(\bar x)$ means that for every graph $G$ and every tuple $\bar v\in V(G)^{|\bar x|}$ we have $G\models \psi^{(r)}(\bar v)$  if and only if $G[N_r(\bar v)]\models \psi^{(r)}(\bar v)$, 
where $G[N_r(\bar v)]$ denotes the subgraph of~$G$ induced by the $r$-neighborhood $N_r(\bar v)$ of $\bar v$. 
The numbers~$r$ and $k$ in the formulas above depend only on the formula~$\varphi$, and furthermore, the Gaifman normal form of any formula $\varphi$ is computable from $\varphi$. 

This translates the model-checking problem to the following algorithmic problem. 
To decide for a graph $G$ and tuple $\bar v\in V(G)^{|\bar x|}$ whether $G\models \varphi(\bar v)$, 
\vspace{1mm}
\begin{enumerate}
\item decide whether $\bar v$ has the local properties described
by $\psi^{(r)}(\bar x)$;
\item decide for each $v\in V(G)$ whether $G\models \chi^{(r)}(v)$;
\item solve each \emph{generalized
independent set} problem described by the basic local formulas \linebreak
$\exists x_1\ldots \exists x_k\big(\bigwedge_{i\neq j}
\dist(x_i,x_j)>2r \wedge \chi^{(r)}(x_i)\big)$, and finally
\item evaluate the Boolean combination of these 
statements that is equivalent to $\varphi$. 
\end{enumerate}  

Note that the generalized independent set problem is also a local problem by the following argument. 
If a graph has many vertices satisfying~$\chi^{(r)}$ that are far away from each other, then we can greedily pick elements for the independent set. 
Otherwise, all elements satisfying~$\chi^{(r)}$ are close to one of the greedily picked vertices, so that testing if there is a different solution becomes a local property. 

Local formulas can be evaluated in bounded-radius neighborhoods of the graph, where the radius depends only on $\phi$. 
Hence, whenever the local neighborhoods in graphs from a class $\Cc$ admit efficient model checking, then one immediately obtains an efficient model checking algorithm for $\Cc$. 
This is e.g.\ the case for planar graphs, as they have locally bounded treewidth, or on classes with locally bounded cliquewidth, such as map graphs. 
The technique based on Gaifman's Theorem was first employed in~\cite{frick2001deciding}. 

In more general graph classes it may be the case that in local neighborhoods we cannot immediately apply one of the known meta theorems. 
However, it may be the case that we can simplify the neighborhoods by the interpretation method, and then apply the locality based method recursively. 
This is the case in nowhere dense and monadically stable classes of graphs (which will be defined in a moment). 
In nowhere dense classes we can simplify neighborhoods by deleting a single vertex and in monadically stable classes we can simplify by \emph{flipping} the edges between two sets of vertices. 
Obviously, the original edge set can be recovered by an interpretation in a coloring of the graph in both cases. 
By appropriately chosen vertex deletions or flips, we have made the formula to be checked more complicated, but the graph to be tested has become simpler. 
We then recurse with localizing the formulas, considering local neighborhoods and deleting a vertex or flipping, until finally we arrive at a single vertex graph, on which the model checking problem is trivial. 
One important requirement is that the recursion terminates in a number of steps that depends only on the input formula $\phi$. 
The formal treatment of the recursive simplification of the graph is captured by games, by the \emph{Splitter game} on nowhere dense classes and the \emph{Flipper game} on monadically stable classes. 
Let us define these classes, as well as the even more general monadically dependent classes, which are conjectured to be the most general hereditary classes of graphs with tractable \FO model checking. 

\subsection{Monadic Stability and Dependence}

Let $\phi(\bar x,\bar y)$ be an \FO-formula. 
Let $\strA$ be a structure and let $A$ be a set of $|\bar x|$-tuples of elements of~$\strA$. 
We call $A$ \emph{shattered} by $\phi$ if there exists a family $\{\bar b_I \mid I\subseteq A\}$ of $|\bar y|$-tuples such that 
\[\strA\models \phi(\bar a,\bar b_I)\Leftrightarrow \bar a\in I\quad \text{for all $\bar a\in A$}.\]

Let $\Cc$ be a class of structures. 
The maximal integer $m$ for which there exists some $A$ of size $m$ in some structure from $\Cc$ shattered by
$\phi(\bar x,\bar y)$ (or $\infty$ if no bound exists) is called the \emph{VC-dimension} of $\phi$ in $\Cc$. 
A class $\Cc$ of structures is \emph{dependent} or \emph{NIP} if for every formula $\phi(\bar x,\bar y)$ has finite VC-dimension in $\Cc$. 

The formula $\phi(\bar x,\bar y)$ has the \emph{$n$-order property} in $\strA$ if there are sequences $\bar a_1,\ldots, \bar a_n$ and $\bar b_1,\ldots, \bar b_n$ of $|\bar x|$- and $|\bar y|$-tuples, respectively, such that 
\[\strA\models \phi(\bar a_i,\bar b_j)\Leftrightarrow i\leq j.\]

The maximal integer $n$ for which there exists $\strA$ in $\Cc$ such that that $\phi$ has the $n$-order property is called the \emph{order-dimension} or \emph{Littlestone-dimension} of $\phi$ in $\Cc$. 
The class $\Cc$ is called \emph{stable} if every formula $\phi(\bar x,\bar y)$ has finite order dimension in $\Cc$. 

The notions of dependence and stability play a key role in classical model theory, see the legendary monograph of Shelah~\cite{shelah1990classification}. 
Obviously, every stable class is also dependent. 
A class is \emph{monadically stable/dependent} if all its monadic lifts are stable/dependent. 
As shown by Baldwin and Shelah~\cite{baldwin1985second} a class $\Cc$ is monadically stable/dependent if and only if all formulas $\phi(x,y)$ with two singleton free variables has finite VC/order-dimension in $\Cc$. 
This yields a very convenient characterization of monadically stable/dependent classes via transductions.
We call the bipartite graph with vertices $a_1,\ldots, a_n$ and $b_1,\ldots, b_n$ and edges $a_ib_j$ for $i\leq j$ a \emph{half-graph} of order $n$.
A class~$\Cc$ is monadically stable/dependent if and only if one cannot transduce all half-graphs/all graphs from $\Cc$. 
An important observation by Adler and Adler~\cite{adler2014interpreting} showed that for monotone classes~$\Cc$ the following are equivalent. 
\begin{enumerate}
    \item $\Cc$ is dependent. \\[-6mm]
    \item $\Cc$ is monadically dependent. \\[-6mm]
    \item $\Cc$ is stable. \\[-6mm]
    \item $\Cc$ is monadically stable.\\[-6mm]
    \item $\Cc$ is nowhere dense.
\end{enumerate}

For monotone classes nowhere denseness constitutes the limit of tractability~\cite{GroheKS17, dvovrak2010deciding, kreutzer2011algorithmic}, and, by the above characterization, one could also say that on monotone classes (monadic) dependence constitutes the limit of tractability. 
As shown by Braunfeld and Laskowski~\cite{braunfeld2022existential} a hereditary class $\Cc$ is dependent if and only if it is monadically dependent and it is stable if and only if it is monadically stable. 
This result further motivates the study of hereditary monadically stable/dependent classes. 

\subsection{Game-Based Decompositions of Local Neighborhoods}

The recursive decomposition of local neighborhoods via games was first introduced in~\cite{GroheKS17}, and could be generalized to monadically stable classes~\cite{dreier2023first, dreier2023first2}. 

In the radius-$r$ Splitter game, two players called \emph{Connector} and \emph{Splitter}, engage on a graph and thereby recursively decompose local neighborhoods. 
Starting with the input graph~$G$, in each of the following rounds, Connector chooses a subgraph of the current game graph of radius at most~$r$ and Splitter deletes a single vertex from this graph. 
The game continues with the resulting graph and terminates when the single-vertex graph is reached. 
A class of graphs is nowhere dense if and only if for every $r$ there exists $\ell$ such that Splitter can win the radius-$r$ Splitter game in $\ell$ rounds~\cite{GroheKS17}. Furthermore, a strategy for splitter can be efficiently computed. Hence, the recursive decomposition can be efficiently computed and terminates after a bounded number of rounds. 

Similarly, in the radius-$r$ Flipper game, two players called \emph{Connector} and \emph{Flipper}, engage on a graph and thereby recursively decompose local neighborhoods. 
Starting with the input graph $G$, in each of the following rounds, Connector chooses a subgraph of the current game graph of radius at most~$r$ and Flipper chooses two sets of vertices $A$ and $B$ and flips the adjacency between the vertices of these two sets. 
The game continues with the resulting graph. Flipper wins once a graph consisting of a single vertex is reached. 
A class of graphs is monadically stable if and only if for every $r$ there exists $\ell(r)$ such that Flipper can win the radius-$r$ Flipper game in $\ell(r)$ rounds~\cite{GajarskyMMOPPSS23}. 
Furthermore, a strategy for flipper can be efficiently computed. 
Again, we get an efficient recursive decomposition of bounded depth. 

\subsection{Model Checking}

A straight-forward implementation of recursive localization (of all $r$-neighborhoods) guided by a game however leads to a branching degree of $n$ and a running time of $\Omega(n^{\ell})$, where~the depth $\ell$ of the recursion grows with~$\phi$. 
It was shown in~\cite{GroheKS17} how to avoid this problem by clustering nearby neighborhoods and handle them together in one recursive call using so-called \emph{sparse $r$-neighborhood covers}. 
An $r$-neighborhood cover with \emph{degree} $d$ and \emph{radius} $s$ of a graph~$G$ is a family~$\Xx$ of subsets of~$V(G)$, called \emph{clusters}, such that the $r$-neighborhood of every vertex is contained in some cluster, every cluster has radius at most $s$, and every vertex appears in at most $d$ clusters.
It was shown in~\cite{GroheKS17} that nowhere dense classes admit $r$-neighborhood covers with radius $2r$ and degree $\Oof(n^\epsilon)$ for any $\epsilon>0$. 
Similarly, it was shown in~\cite{dreier2023first2} that monadically stable classes admit $r$-neighborhood covers with radius $4r$ and degree $\Oof(n^\epsilon)$ for any $\epsilon>0$. 
By scaling $\epsilon$ appropriately the recursive data structure can be constructed in the desired fpt running time. 

Some care has to be taken that the quantifier rank of the formulas does not grow when localizing in each recursive step. 
This technical problem was addressed by considering a rank-preserving normal form in~\cite{GroheKS17}. 
This approach could be much simplified by a nice game based argument in~\cite{dreier2023first}. We obtain the following theorems. 

\begin{theorem}[\cite{GroheKS17}]
Let $\Cc$ be a nowhere dense class. 
    Then \emph{MC}$(\FO,\Cc)$ is fixed-parameter tractable. 
\end{theorem}

\begin{theorem}[\cite{dreier2023first,dreier2023first2}]
    Let $\Cc$ be a monadically stable class. 
    Then \emph{MC}$(\FO,\Cc)$ is fixed-parameter tractable.
\end{theorem}

As mentioned in the introduction, it is one of the main questions in the area whether these results can be extended to monadically dependent classes. 
If based on the same approach as for nowhere dense and monadically stable classes, one would have to devise a game that not necessarily has bounded length, but that, combined with sparse neighborhood coves, if they exist for monadically dependent classes, leads to an fpt bounded size recursive data structure. 

\section{Twinwidth}

\emph{Twinwidth} was recently introduced by
Bonnet, Kim, Thomass\'e and Watrigant~\cite{bonnet2021twin} and immediately had a major impact in (algorithmic) graph theory. 
Many well-studied classes of graphs have bounded twinwidth, e.g.\ planar graphs, and more generally, any class of graphs excluding a fixed minor, cographs, and more generally, any class of bounded clique-width, and many more. 
Most imporantly in our context, the property of having bounded twinwidth is preserved unter \FO-transductions, hence, classes with bounded twinwidth are monadically dependent. 
In fact, a class of ordered structures has bounded twinwidth if and only if it is monadically dependent~\cite{bonnet2024twin}. 
\FO model checking is fixed-parameter tractable on every class of bounded twinwidth assuming a contraction sequence is given together with the input~\cite{bonnet2021twin}. On ordered structures a contraction sequence can be computed efficiently, leading to efficient model checking on ordered classes of bounded twinwidth~\cite{bonnet2024twin}, supporting the conjecture that \FO model checking on hereditary classes is tractable if and only if the class is monadically dependent. 
We give a brief sketch of model checking on classes with bounded twinwidth, following the presentation of~\cite{GajarskyPPT22}. 

The definition of twinwidth is based on \emph{contraction sequences}. 
Let $G$ be a graph on $n$ vertices.
A \emph{contraction sequence} for $G$
is a sequence $\Pp_1,\ldots,\Pp_n$
of partitions of the vertex set of $G$ such that:
\vspace{1mm}
\begin{itemize} 
    \item $\Pp_1$ is the partition into singletons;
     \item $\Pp_n$ is the partition with one part;
    \item for each $t\in [n]$, $t>1$, $\Pp_{t}$ is obtained from $\Pp_{t-1}$ by taking some two parts $A,B\in \Pp_{t-1}$ and {\em{contracting}} them: replacing them with a single part $A\cup B\in \Pp_{t}$.
\end{itemize}

A pair of disjoint vertex subsets $A,B\subset V(G)$ is {\em{complete}} if every vertex of $A$ is adjacent to every vertex of $B$, and {\em{anti-complete}} if there is no edge with one endpoint in $A$ and the other one in~$B$. 
The pair $A,B$ is {\em{pure}} if it is complete or anti-complete, and {\em{impure}} otherwise.

A {\em{trigraph}} is a structure with two types of undirected edges, \emph{red} and \emph{black edges}. 
Given a partition of $\Pp$, we define the {\em{quotient trigraph}} $G/\Pp$ as the trigraph on vertex set $\Pp$, where two parts $A,B\in \Pp$ are connected by a black edge if the pair $A,B$ is complete in $G$, 
non-adjacent if the pair is anti-complete, 
and connected by a red edge if $A,B$ is impure. 
For a trigraph $H$, its {\em{impurity graph}} is the graph on vertex set $H$ where two vertices $u,v\in V(H)$ are considered adjacent if they are impure towards each other in $H$, that is, the subgraph on $V(H)$ induced by the red edges of $H$. 

The {\em{width}} of the contraction sequence $\Pp_1,\ldots,\Pp_n$ is the maximum degree in the impurity graphs of $G/\Pp_t$ over all times $t\in [n]$. The {\em{twin-width}} of $G$ is the minimum possible width of a contraction sequence of $G$.

The model checking result can be presented using the notion of {\em local types}. 
Very roughly, the $k$-local type of a part refers to all formulas (up to quantifier rank $k$) that the $2^k$-neighborhood of the part satisfies, where we consider distances in the impurity graphs. 

Initially, the first partition of the sequence only contains singletons and no part is impure with another. So the local types (for a fixed value of $k$) can be computed in constant time for each part.

Then, and since the provided sequence has width at most $d$, the $k$-neighborhood of a part only contains $d^k$ other parts. Furthermore, when two parts are merged according to the sequence, this only impacts the $k$-neighborhood of a constant number of parts. These types can be recomputed efficiently.

Finally, the local type of the last partition, which only contains one part: the entire graph, provides the type of the graph and informs us whether the original desired property is satisfied.

\subsection{Around Twinwidth}
The method described above, gives rise to various questions. Are there other suitable width parameters for the contraction sequence? Do we need the sequence to end on a single part, or could there be other stopping points?

Some of these directions have been considered by Bonnet et al.~\cite{DBLP:conf/soda/BonnetKRT22}. First, by changing the notion of width for the contraction sequence, one can capture the notions of rank width and linear rank width by bounding the size of connected components in the impurity graph, or the total number of impure connections. 

On the direction of contraction sequences that stop before the single part partition, Bonnet et al.~\cite{DBLP:conf/soda/BonnetKRT22} also proved that if the connections (both pure and impure) of a partition results in a graph of bounded degree, or a graph of bounded expansion, one could still derive information on the type of the original graph. 
Proving fixed parameter tractability of \FO model checking on classes of graphs that can contract to a class with bounded degree, and fixed parameter tractability of $\exists\FO$ model checking on classes of graphs that can contract to a class with bounded expansion. 
Where $\exists\FO$ is the existential fragment of first order logic, which prohibits the use of universal quantification, as well as negations.

\bibliographystyle{plain}
\bibliography{ref}

\begin{thebibliography}{100}

\bibitem{adler2014interpreting}
Hans Adler and Isolde Adler.
\newblock Interpreting nowhere dense graph classes as a classical notion of model theory.
\newblock {\em European Journal of Combinatorics}, 36:322--330, 2014.

\bibitem{baldwin1985second}
John~T Baldwin and Saharon Shelah.
\newblock Second-order quantifiers and the complexity of theories.
\newblock {\em Notre Dame Journal of Formal Logic}, 26(3):229--303, 1985.

\bibitem{barrington1990uniformity}
David A~Mix Barrington, Neil Immerman, and Howard Straubing.
\newblock On uniformity within {NC1}.
\newblock {\em Journal of Computer and System Sciences}, 41(3):274--306, 1990.

\bibitem{benedikt2005expressive}
Michael Benedikt and H~Jerome Keisler.
\newblock Expressive power of unary counters.
\newblock {\em Structures in Logic and Computer Science: A Selection of Essays in Honor of A. Ehrenfeucht}, pages 34--50, 2005.

\bibitem{bojanczyk2021separator}
Mikolaj Bojanczyk.
\newblock Separator logic and star-free expressions for graphs.
\newblock {\em arXiv preprint arXiv:2107.13953}, 2021.

\bibitem{bonnet2022twinwin}
{\'{E}}douard Bonnet, Dibyayan Chakraborty, Eun~Jung Kim, Noleen K{\"{o}}hler, Raul Lopes, and St{\'{e}}phan Thomass{\'{e}}.
\newblock Twin-width {VIII:} delineation and {Win}-{Wins}.
\newblock In {\em 17th International Symposium on Parameterized and Exact Computation, {IPEC} 2022}, volume 249 of {\em LIPIcs}, pages 9:1--9:18. Schloss Dagstuhl - Leibniz-Zentrum f{\"{u}}r Informatik, 2022.

\bibitem{bonnet2022model}
{\'E}douard Bonnet, Jan Dreier, Jakub Gajarsk{\`y}, Stephan Kreutzer, Nikolas M{\"a}hlmann, Pierre Simon, and Szymon Toru{\'n}czyk.
\newblock Model checking on interpretations of classes of bounded local cliquewidth.
\newblock In {\em Proceedings of the 37th Annual ACM/IEEE Symposium on Logic in Computer Science}, pages 1--13, 2022.

\bibitem{bonnet2021twin2}
{\'E}douard Bonnet, Colin Geniet, Eun~Jung Kim, St{\'e}phan Thomass{\'e}, and R{\'e}mi Watrigant.
\newblock Twin-width {II}: small classes.
\newblock In {\em Proceedings of the 2021 ACM-SIAM Symposium on Discrete Algorithms (SODA)}, pages 1977--1996. SIAM, 2021.

\bibitem{BonnetG0TW21}
{\'{E}}douard Bonnet, Colin Geniet, Eun~Jung Kim, St{\'{e}}phan Thomass{\'{e}}, and R{\'{e}}mi Watrigant.
\newblock Twin-width {III:} max independent set, min dominating set, and coloring.
\newblock In {\em 48th International Colloquium on Automata, Languages, and Programming, {ICALP} 2021}, volume 198 of {\em LIPIcs}, pages 35:1--35:20. Schloss Dagstuhl - Leibniz-Zentrum f{\"{u}}r Informatik, 2021.

\bibitem{bonnet2024twin}
{\'E}douard Bonnet, Ugo Giocanti, Patrice~Ossona de~Mendez, Pierre Simon, St{\'e}phan Thomass{\'e}, and Szymon Toru{\'n}czyk.
\newblock Twin-width {IV}: ordered graphs and matrices.
\newblock {\em Journal of the ACM}, 71(3):1--45, 2024.

\bibitem{BonnetGMT23}
{\'{E}}douard Bonnet, Ugo Giocanti, Patrice~Ossona de~Mendez, and St{\'{e}}phan Thomass{\'{e}}.
\newblock Twin-width {V:} linear minors, modular counting, and matrix multiplication.
\newblock In {\em 40th International Symposium on Theoretical Aspects of Computer Science, {STACS} 2023}, volume 254 of {\em LIPIcs}, pages 15:1--15:16. Schloss Dagstuhl - Leibniz-Zentrum f{\"{u}}r Informatik, 2023.

\bibitem{DBLP:conf/soda/BonnetKRT22}
{\'{E}}douard Bonnet, Eun~Jung Kim, Amadeus Reinald, and St{\'{e}}phan Thomass{\'{e}}.
\newblock Twin-width {VI:} the lens of contraction sequences.
\newblock In {\em {SODA}}, pages 1036--1056. {SIAM}, 2022.

\bibitem{bonnet2022twin}
{\'E}douard Bonnet, Eun~Jung Kim, Amadeus Reinald, St{\'e}phan Thomass{\'e}, and R{\'e}mi Watrigant.
\newblock Twin-width and polynomial kernels.
\newblock {\em Algorithmica}, 84(11):3300--3337, 2022.

\bibitem{bonnet2021twin}
{\'E}douard Bonnet, Eun~Jung Kim, St{\'e}phan Thomass{\'e}, and R{\'e}mi Watrigant.
\newblock Twin-width {I}: tractable {FO} model checking.
\newblock {\em ACM Journal of the ACM (JACM)}, 69(1):1--46, 2021.

\bibitem{bonnet2024twinperm}
{\'E}douard Bonnet, Jaroslav Ne{\v{s}}et{\v{r}}il, Patrice~Ossona de~Mendez, Sebastian Siebertz, and St{\'e}phan Thomass{\'e}.
\newblock Twin-width and permutations.
\newblock {\em Logical Methods in Computer Science}, 20, 2024.

\bibitem{braunfeld2022existential}
Samuel Braunfeld and Michael~C Laskowski.
\newblock Existential characterizations of monadic {NIP}.
\newblock {\em arXiv preprint arXiv:2209.05120}, 2022.

\bibitem{braunfeld2022decomposition}
Samuel Braunfeld, Jaroslav Ne{\v{s}}et{\v{r}}il, Patrice~Ossona de~Mendez, and Sebastian Siebertz.
\newblock Decomposition horizons: from graph sparsity to model-theoretic dividing lines.
\newblock {\em arXiv preprint arXiv:2209.11229}, 2022.

\bibitem{braunfeld2024decomposition}
Samuel Braunfeld, Jaroslav Ne{\v{s}}et{\v{r}}il, Patrice~Ossona de~Mendez, and Sebastian Siebertz.
\newblock Decomposition horizons and a characterization of stable hereditary classes of graphs.
\newblock {\em arXiv preprint arXiv:2209.11229}, 2024.

\bibitem{chandra1982structure}
Ashok Chandra and David Harel.
\newblock Structure and complexity of relational queries.
\newblock {\em Journal of Computer and system Sciences}, 25(1):99--128, 1982.

\bibitem{chandra1977optimal}
Ashok~K Chandra and Philip~M Merlin.
\newblock Optimal implementation of conjunctive queries in relational data bases.
\newblock In {\em Proceedings of the ninth annual ACM symposium on Theory of computing}, pages 77--90, 1977.

\bibitem{chen2006strong}
Jianer Chen, Xiuzhen Huang, Iyad~A Kanj, and Ge~Xia.
\newblock Strong computational lower bounds via parameterized complexity.
\newblock {\em Journal of Computer and System Sciences}, 72(8):1346--1367, 2006.

\bibitem{chitnis2016designing}
Rajesh Chitnis, Marek Cygan, MohammadTaghi Hajiaghayi, Marcin Pilipczuk, and Micha{\l} Pilipczuk.
\newblock Designing fpt algorithms for cut problems using randomized contractions.
\newblock {\em SIAM Journal on Computing}, 45(4):1171--1229, 2016.

\bibitem{codd1970relational}
Edgar~F Codd.
\newblock A relational model of data for large shared data banks.
\newblock {\em Communications of the ACM}, 13(6):377--387, 1970.

\bibitem{colcombet2007combinatorial}
Thomas Colcombet.
\newblock A combinatorial theorem for trees: applications to monadic logic and infinite structures.
\newblock In {\em Automata, Languages and Programming: 34th International Colloquium, ICALP 2007}, pages 901--912. Springer, 2007.

\bibitem{courcelle1990graph}
Bruno Courcelle.
\newblock Graph rewriting: An algebraic and logic approach.
\newblock In {\em Formal models and semantics}, pages 193--242. Elsevier, 1990.

\bibitem{courcelle1990monadic}
Bruno Courcelle.
\newblock The monadic second-order logic of graphs. {I}. {Recognizable} sets of finite graphs.
\newblock {\em Information and computation}, 85(1):12--75, 1990.

\bibitem{courcelle1992monadic}
Bruno Courcelle.
\newblock The monadic second-order logic of graphs {VII}: Graphs as relational structures.
\newblock {\em Theoretical Computer Science}, 101(1):3--33, 1992.

\bibitem{courcelle2000linear}
Bruno Courcelle, Johann~A Makowsky, and Udi Rotics.
\newblock Linear time solvable optimization problems on graphs of bounded clique-width.
\newblock {\em Theory of Computing Systems}, 33(2):125--150, 2000.

\bibitem{courcelle2000upper}
Bruno Courcelle and Stephan Olariu.
\newblock Upper bounds to the clique width of graphs.
\newblock {\em Discrete Applied Mathematics}, 101(1-3):77--114, 2000.

\bibitem{courcelle2007vertex}
Bruno Courcelle and Sang-il Oum.
\newblock Vertex-minors, monadic second-order logic, and a conjecture by {S}eese.
\newblock {\em Journal of Combinatorial Theory, Series B}, 97(1):91--126, 2007.

\bibitem{cygan2015parameterized}
Marek Cygan, Fedor~V Fomin, {\L}ukasz Kowalik, Daniel Lokshtanov, D{\'a}niel Marx, Marcin Pilipczuk, Micha{\l} Pilipczuk, and Saket Saurabh.
\newblock {\em Parameterized algorithms}, volume~5.
\newblock Springer, 2015.

\bibitem{cygan2020randomized}
Marek Cygan, Pawe{\l} Komosa, Daniel Lokshtanov, Marcin Pilipczuk, Micha{\l} Pilipczuk, Saket Saurabh, and Magnus Wahlstr{\"o}m.
\newblock Randomized contractions meet lean decompositions.
\newblock {\em ACM Transactions on Algorithms (TALG)}, 17(1):1--30, 2020.

\bibitem{cygan2014minimum}
Marek Cygan, Daniel Lokshtanov, Marcin Pilipczuk, Micha{\l} Pilipczuk, and Saket Saurabh.
\newblock Minimum bisection is fixed parameter tractable.
\newblock In {\em Proceedings of the forty-sixth annual ACM symposium on Theory of computing}, pages 323--332, 2014.

\bibitem{cygan2019minimum}
Marek Cygan, Daniel Lokshtanov, Marcin Pilipczuk, Micha{\l} Pilipczuk, and Saket Saurabh.
\newblock Minimum bisection is fixed-parameter tractable.
\newblock {\em SIAM Journal on Computing}, 48(2):417--450, 2019.

\bibitem{dawar2007locally}
Anuj Dawar, Martin Grohe, and Stephan Kreutzer.
\newblock Locally excluding a minor.
\newblock In {\em LICS 2007}, pages 270--279. IEEE, 2007.

\bibitem{doner1970tree}
John Doner.
\newblock Tree acceptors and some of their applications.
\newblock {\em Journal of Computer and System Sciences}, 4(5):406--451, 1970.

\bibitem{downey1995fixed}
Rod~G Downey and Michael~R Fellows.
\newblock Fixed-parameter tractability and completeness {I}: Basic results.
\newblock {\em SIAM Journal on computing}, 24(4):873--921, 1995.

\bibitem{DowneyF99}
Rodney~G. Downey and Michael~R. Fellows.
\newblock {\em Parameterized Complexity}.
\newblock Monographs in Computer Science. Springer, 1999.

\bibitem{downey2013fundamentals}
Rodney~G Downey, Michael~R Fellows, et~al.
\newblock {\em Fundamentals of parameterized complexity}, volume~4.
\newblock Springer, 2013.

\bibitem{dreier2021lacon}
Jan Dreier.
\newblock Lacon-and shrub-decompositions: A new characterization of first-order transductions of bounded expansion classes.
\newblock In {\em 2021 36th Annual ACM/IEEE Symposium on Logic in Computer Science (LICS)}, pages 1--13. IEEE, 2021.

\bibitem{dreier2023first2}
Jan Dreier, Ioannis Eleftheriadis, Nikolas M{\"a}hlmann, Rose McCarty, Micha{\l} Pilipczuk, and Szymon Toru{\'n}czyk.
\newblock First-order model checking on monadically stable graph classes.
\newblock {\em arXiv preprint arXiv:2311.18740}, 2023.

\bibitem{dreier2022treelike}
Jan Dreier, Jakub Gajarsk{\`y}, Sandra Kiefer, Micha{\l} Pilipczuk, and Szymon Toru{\'n}czyk.
\newblock Treelike decompositions for transductions of sparse graphs.
\newblock In {\em Proceedings of the 37th Annual ACM/IEEE Symposium on Logic in Computer Science}, pages 1--14, 2022.

\bibitem{DreierMMSV22}
Jan Dreier, Nikolas M{\"{a}}hlmann, Amer~E. Mouawad, Sebastian Siebertz, and Alexandre Vigny.
\newblock Combinatorial and algorithmic aspects of monadic stability.
\newblock In {\em 33rd International Symposium on Algorithms and Computation, {ISAAC} 2022}, volume 248 of {\em LIPIcs}, pages 11:1--11:17. Schloss Dagstuhl - Leibniz-Zentrum f{\"{u}}r Informatik, 2022.

\bibitem{dreier2023first}
Jan Dreier, Nikolas M{\"a}hlmann, and Sebastian Siebertz.
\newblock First-order model checking on structurally sparse graph classes.
\newblock In {\em Proceedings of the 55th Annual ACM Symposium on Theory of Computing}, pages 567--580, 2023.

\bibitem{dreier2023indiscernibles}
Jan Dreier, Nikolas M{\"a}hlmann, Sebastian Siebertz, and Szymon Toru{\'n}czyk.
\newblock Indiscernibles and flatness in monadically stable and monadically nip classes.
\newblock In {\em 50th International Colloquium on Automata, Languages, and Programming (ICALP 2023)}. Schloss-Dagstuhl-Leibniz Zentrum f{\"u}r Informatik, 2023.

\bibitem{dreier2024flip}
Jan Dreier, Nikolas M{\"a}hlmann, and Szymon Toru{\'n}czyk.
\newblock Flip-breakability: A combinatorial dichotomy for monadically dependent graph classes.
\newblock In {\em Proceedings of the 56th Annual ACM Symposium on Theory of Computing}, pages 1550--1560, 2024.

\bibitem{dvovrak2010deciding}
Zden{\v{e}}k Dvo{\v{r}}{\'a}k, Daniel Kr{\'a}l, and Robin Thomas.
\newblock Deciding first-order properties for sparse graphs.
\newblock In {\em 2010 IEEE 51st Annual Symposium on Foundations of Computer Science}, pages 133--142. IEEE, 2010.

\bibitem{dvovrak2013testing}
Zden{\v{e}}k Dvo{\v{r}}{\'a}k, Daniel Kr{\'a}l, and Robin Thomas.
\newblock Testing first-order properties for subclasses of sparse graphs.
\newblock {\em Journal of the ACM (JACM)}, 60(5):36, 2013.

\bibitem{eflum}
Heinz{-}Dieter Ebbinghaus and J{\"{o}}rg Flum.
\newblock {\em Finite model theory}.
\newblock Perspectives in Mathematical Logic. Springer, 1995.

\bibitem{Ebbinghaus94}
Heinz{-}Dieter Ebbinghaus, J{\"{o}}rg Flum, and Wolfgang Thomas.
\newblock {\em Mathematical logic {(2.} ed.)}.
\newblock Undergraduate texts in mathematics. Springer, 1994.

\bibitem{eickmeyer2020model}
Kord Eickmeyer, Jan van~den Heuvel, Ken-ichi Kawarabayashi, Stephan Kreutzer, Patrice Ossona~De Mendez, Micha{\l} Pilipczuk, Daniel~A Quiroz, Roman Rabinovich, and Sebastian Siebertz.
\newblock Model-checking on ordered structures.
\newblock {\em ACM Transactions on Computational Logic (TOCL)}, 21(2):1--28, 2020.

\bibitem{etessami1997counting}
Kousha Etessami.
\newblock Counting quantifiers, successor relations, and logarithmic space.
\newblock {\em Journal of Computer and System Sciences}, 54(3):400--411, 1997.

\bibitem{fagin1974generalized}
Ronald Fagin.
\newblock Generalized first-order spectra and polynomial-time recognizable sets.
\newblock {\em Complexity of computation}, 7:43--73, 1974.

\bibitem{fagin1975monadic}
Ronald Fagin.
\newblock Monadic generalized spectra.
\newblock {\em Math. Log. Q.}, 21(1):89--96, 1975.

\bibitem{feferman1967first}
Solomon Feferman and Robert~L Vaught.
\newblock The first order properties of products of algebraic systems.
\newblock {\em Journal of Symbolic Logic}, 32(2), 1967.

\bibitem{flum2001fixed}
J{\"o}rg Flum and Martin Grohe.
\newblock Fixed-parameter tractability, definability, and model-checking.
\newblock {\em SIAM Journal on Computing}, 31(1):113--145, 2001.

\bibitem{FlumG06}
J{\"{o}}rg Flum and Martin Grohe.
\newblock {\em Parameterized Complexity Theory}.
\newblock Texts in Theoretical Computer Science. An {EATCS} Series. Springer, 2006.

\bibitem{fomin2023compound}
Fedor~V Fomin, Petr~A Golovach, Ignasi Sau, Giannos Stamoulis, and Dimitrios~M Thilikos.
\newblock Compound logics for modification problems.
\newblock {\em ACM Transactions on Computational Logic}, 2023.

\bibitem{fomin2023algorithmic}
Fedor~V Fomin, Petr~A Golovach, Giannos Stamoulis, and Dimitrios~M Thilikos.
\newblock An algorithmic meta-theorem for graph modification to planarity and fol.
\newblock {\em ACM Transactions on Computation Theory}, 14(3-4):1--29, 2023.

\bibitem{frick2001deciding}
Markus Frick and Martin Grohe.
\newblock Deciding first-order properties of locally tree-decomposable structures.
\newblock {\em Journal of the ACM (JACM)}, 48(6):1184--1206, 2001.

\bibitem{gaifman1982local}
Haim Gaifman.
\newblock On local and non-local properties.
\newblock In {\em Studies in Logic and the Foundations of Mathematics}, volume 107, pages 105--135. Elsevier, 1982.

\bibitem{gajarsky2020new}
Jakub Gajarsk{\`y}, Petr Hlin{\v{e}}n{\`y}, Jan Obdr{\v{z}}{\'a}lek, Daniel Lokshtanov, and M~Sridharan Ramanujan.
\newblock A new perspective on {FO} model checking of dense graph classes.
\newblock {\em ACM Transactions on Computational Logic (TOCL)}, 21(4):1--23, 2020.

\bibitem{gajarsky2018recovering}
Jakub Gajarsky and Daniel Kr{\'a}l'.
\newblock Recovering sparse graphs.
\newblock {\em Leibniz International Proceedings in Informatics (LIPIcs)}, 117:29, 2018.

\bibitem{gajarsky2020first}
Jakub Gajarsk{\`y}, Stephan Kreutzer, Jaroslav Ne{\v{s}}et{\v{r}}il, Patrice Ossona~De Mendez, Micha{\l} Pilipczuk, Sebastian Siebertz, and Szymon Toru{\'n}czyk.
\newblock First-order interpretations of bounded expansion classes.
\newblock {\em ACM Transactions on Computational Logic (TOCL)}, 21(4):1--41, 2020.

\bibitem{GajarskyMMOPPSS23}
Jakub Gajarsk{\'{y}}, Nikolas M{\"{a}}hlmann, Rose McCarty, Pierre Ohlmann, Michal Pilipczuk, Wojciech Przybyszewski, Sebastian Siebertz, Marek Sokolowski, and Szymon Torunczyk.
\newblock Flipper games for monadically stable graph classes.
\newblock In {\em 50th International Colloquium on Automata, Languages, and Programming, {ICALP} 2023}, volume 261 of {\em LIPIcs}, pages 128:1--128:16. Schloss Dagstuhl - Leibniz-Zentrum f{\"{u}}r Informatik, 2023.

\bibitem{GajarskyPPT22}
Jakub Gajarsk{\'{y}}, Michal Pilipczuk, Wojciech Przybyszewski, and Szymon Torunczyk.
\newblock Twin-width and types.
\newblock In {\em 49th International Colloquium on Automata, Languages, and Programming, {ICALP} 2022}, volume 229 of {\em LIPIcs}, pages 123:1--123:21. Schloss Dagstuhl - Leibniz-Zentrum f{\"{u}}r Informatik, 2022.

\bibitem{gajarsky2022stable}
Jakub Gajarsk{\`y}, Micha{\l} Pilipczuk, and Szymon Toru{\'n}czyk.
\newblock Stable graphs of bounded twin-width.
\newblock In {\em Proceedings of the 37th Annual ACM/IEEE Symposium on Logic in Computer Science}, pages 1--12, 2022.

\bibitem{ganian2014lower}
Robert Ganian, Petr Hlin{\v{e}}n{\`y}, Alexander Langer, Jan Obdr{\v{z}}{\'a}lek, Peter Rossmanith, and Somnath Sikdar.
\newblock Lower bounds on the complexity of {MSO}$_1$ model-checking.
\newblock {\em Journal of Computer and System Sciences}, 80(1):180--194, 2014.

\bibitem{ganian2019shrub}
Robert Ganian, Petr Hlin{\v{e}}n{\`y}, Jaroslav Ne{\v{s}}et{\v{r}}il, Jan Obdr{\v{z}}{\'a}lek, and Patrice~Ossona De~Mendez.
\newblock Shrub-depth: Capturing height of dense graphs.
\newblock {\em Logical Methods in Computer Science}, 15, 2019.

\bibitem{ganian2012trees}
Robert Ganian, Petr Hlin{\v{e}}n{\`y}, Jaroslav Ne{\v{s}}et{\v{r}}il, Jan Obdr{\v{z}}{\'a}lek, Patrice Ossona~de Mendez, and Reshma Ramadurai.
\newblock When trees grow low: Shrubs and fast {MSO}$_1$.
\newblock In {\em Mathematical Foundations of Computer Science 2012: 37th International Symposium, MFCS 2012, Bratislava, Slovakia, August 27-31, 2012. Proceedings 37}, pages 419--430. Springer, 2012.

\bibitem{GanzowR08}
Tobias Ganzow and Sasha Rubin.
\newblock Order-invariant {MSO} is stronger than counting {MSO} in the finite.
\newblock In Susanne Albers and Pascal Weil, editors, {\em 25th Annual Symposium on Theoretical Aspects of Computer Science, {STACS} 2008}, volume~1 of {\em LIPIcs}, pages 313--324. Schloss Dagstuhl - Leibniz-Zentrum f{\"{u}}r Informatik, Germany, 2008.

\bibitem{geniet2024twin}
Colin Geniet.
\newblock {\em Twin-Width, logical and combinatorial characterisations}.
\newblock PhD thesis, Ecole normale sup{\'e}rieure de lyon-ENS LYON, 2024.

\bibitem{golovach2023model}
Petr~A Golovach, Giannos Stamoulis, and Dimitrios~M Thilikos.
\newblock Model-checking for first-order logic with disjoint paths predicates in proper minor-closed graph classes.
\newblock In {\em Proceedings of the 2023 Annual ACM-SIAM Symposium on Discrete Algorithms (SODA)}, pages 3684--3699. SIAM, 2023.

\bibitem{gradel2007finite}
Erich Gr{\"a}del, Phokion~G Kolaitis, Leonid Libkin, Maarten Marx, Joel Spencer, Moshe~Y Vardi, Yde Venema, Scott Weinstein, et~al.
\newblock {\em Finite Model Theory and its applications}.
\newblock Springer, 2007.

\bibitem{GroblerJMSV24}
Mario Grobler, Yiting Jiang, Patrice~Ossona de~Mendez, Sebastian Siebertz, and Alexandre Vigny.
\newblock Discrepancy and sparsity.
\newblock {\em J. Comb. Theory {B}}, 169:96--133, 2024.

\bibitem{grohe2008logic}
Martin Grohe.
\newblock Logic, graphs, and algorithms.
\newblock {\em Logic and automata}, 2:357--422, 2008.

\bibitem{grohe2011finding}
Martin Grohe, Ken-ichi Kawarabayashi, D{\'a}niel Marx, and Paul Wollan.
\newblock Finding topological subgraphs is fixed-parameter tractable.
\newblock In {\em Proceedings of the forty-third annual ACM symposium on Theory of computing}, pages 479--488, 2011.

\bibitem{grohe2009methods}
Martin Grohe and Stephan Kreutzer.
\newblock Methods for algorithmic meta theorems.
\newblock {\em AMS-ASL Joint Special Session}, 558:181--206, 2009.

\bibitem{GroheKS17}
Martin Grohe, Stephan Kreutzer, and Sebastian Siebertz.
\newblock Deciding first-order properties of nowhere dense graphs.
\newblock {\em J. {ACM}}, 64(3):17:1--17:32, 2017.

\bibitem{grohe2018first}
Martin Grohe and Nicole Schweikardt.
\newblock First-order query evaluation with cardinality conditions.
\newblock In {\em Proceedings of the 37th ACM SIGMOD-SIGACT-SIGAI Symposium on Principles of Database Systems}, pages 253--266, 2018.

\bibitem{gurevich1985logic}
Yuri Gurevich.
\newblock Logic and the challenge of computer science.
\newblock Technical report, 1985.

\bibitem{Hodges93}
Wilfrid Hodges.
\newblock {\em Model theory}, volume~42 of {\em Encyclopedia of mathematics and its applications}.
\newblock Cambridge University Press, 1993.

\bibitem{immerman1982upper}
Neil Immerman.
\newblock Upper and lower bounds for first order expressibility.
\newblock {\em Journal of Computer and System Sciences}, 25(1):76--98, 1982.

\bibitem{Immerman87}
Neil Immerman.
\newblock Languages that capture complexity classes.
\newblock {\em {SIAM} J. Comput.}, 16(4):760--778, 1987.

\bibitem{immerman1998descriptive}
Neil Immerman.
\newblock {\em Descriptive complexity}.
\newblock Springer Science \& Business Media, 1998.

\bibitem{jiang2020regular}
Yiting Jiang, Jaroslav Ne{\v{s}}et{\v{r}}il, Patrice Ossona~de Mendez, and Sebastian Siebertz.
\newblock Regular partitions of gentle graphs.
\newblock {\em Acta Mathematica Hungarica}, 161(2):719--755, 2020.

\bibitem{kawarabayashi2011minimum}
Ken-ichi Kawarabayashi and Mikkel Thorup.
\newblock The minimum k-way cut of bounded size is fixed-parameter tractable.
\newblock In {\em 2011 IEEE 52nd Annual Symposium on Foundations of Computer Science}, pages 160--169. IEEE, 2011.

\bibitem{kreutzer2011algorithmic}
Stephan Kreutzer.
\newblock Algorithmic meta-theorems.
\newblock {\em Finite and algorithmic model theory}, 379:177--270, 2011.

\bibitem{kreutzer2012parameterized}
Stephan Kreutzer.
\newblock On the parameterized intractability of monadic second-order logic.
\newblock {\em Logical Methods in Computer Science}, 8, 2012.

\bibitem{kreutzer2010lower}
Stephan Kreutzer and Siamak Tazari.
\newblock Lower bounds for the complexity of monadic second-order logic.
\newblock In {\em 2010 25th Annual IEEE Symposium on Logic in Computer Science}, pages 189--198. IEEE, 2010.

\bibitem{kreutzer2010brambles}
Stephan Kreutzer and Siamak Tazari.
\newblock On brambles, grid-like minors, and parameterized intractability of monadic second-order logic.
\newblock In {\em Proceedings of the twenty-first annual ACM-SIAM symposium on Discrete Algorithms}, pages 354--364. SIAM, 2010.

\bibitem{kuske2017first}
Dietrich Kuske and Nicole Schweikardt.
\newblock First-order logic with counting.
\newblock In {\em 2017 32nd Annual ACM/IEEE Symposium on Logic in Computer Science (LICS)}, pages 1--12. IEEE, 2017.

\bibitem{kuske2018gaifman}
Dietrich Kuske and Nicole Schweikardt.
\newblock Gaifman normal forms for counting extensions of first-order logic.
\newblock In {\em 45th International Colloquium on Automata, Languages, and Programming (ICALP 2018)}. Schloss Dagstuhl-Leibniz-Zentrum fuer Informatik, 2018.

\bibitem{libkin2004elements}
Leonid Libkin.
\newblock {\em Elements of finite model theory}, volume~41.
\newblock Springer, 2004.

\bibitem{Libkin04}
Leonid Libkin.
\newblock {\em Elements of Finite Model Theory}.
\newblock Texts in Theoretical Computer Science. An {EATCS} Series. Springer, 2004.

\bibitem{LokshtanovR0Z18}
Daniel Lokshtanov, M.~S. Ramanujan, Saket Saurabh, and Meirav Zehavi.
\newblock Reducing {CMSO} model checking to highly connected graphs.
\newblock In {\em 45th International Colloquium on Automata, Languages, and Programming, {ICALP} 2018}, volume 107 of {\em LIPIcs}, pages 135:1--135:14. Schloss Dagstuhl - Leibniz-Zentrum f{\"{u}}r Informatik, 2018.

\bibitem{mahlmann2024monadically}
Nikolas M{\"a}hlmann.
\newblock {\em Monadically stable and monadically dependent graph classes: characterizations and algorithmic meta-theorems}.
\newblock PhD thesis, Universit{\"a}t Bremen, 2024.

\bibitem{makowsky2003tree}
Johann~A Makowsky and JP~Marino.
\newblock Tree-width and the monadic quantifier hierarchy.
\newblock {\em Theoretical computer science}, 303(1):157--170, 2003.

\bibitem{malliaris2014regularity}
Maryanthe Malliaris and Saharon Shelah.
\newblock Regularity lemmas for stable graphs.
\newblock {\em Transactions of the American Mathematical Society}, 366(3):1551--1585, 2014.

\bibitem{nevsetvril2016structural}
Jaroslav Ne{\v{s}}et{\v{r}}il and P~Ossona de~Mendez.
\newblock Structural sparsity.
\newblock {\em Russian Mathematical Surveys}, 71(1):79, 2016.

\bibitem{nevsetvril2011nowhere}
Jaroslav Ne{\v{s}}et{\v{r}}il and Patrice~Ossona De~Mendez.
\newblock On nowhere dense graphs.
\newblock {\em European Journal of Combinatorics}, 32(4):600--617, 2011.

\bibitem{nevsetvril2021rankwidth}
Jaroslav Ne{\v{s}}et{\v{r}}il, Patrice Ossona~de Mendez, Micha{\l} Pilipczuk, Roman Rabinovich, and Sebastian Siebertz.
\newblock Rankwidth meets stability.
\newblock In {\em Proceedings of the 2021 ACM-SIAM Symposium on Discrete Algorithms (SODA)}, pages 2014--2033. SIAM, 2021.

\bibitem{nevsetvril2020linear}
Jaroslav Ne{\v{s}}et{\v{r}}il, Roman Rabinovich, Patrice~Ossona de~Mendez, and Sebastian Siebertz.
\newblock Linear rankwidth meets stability.
\newblock In {\em Proceedings of the Fourteenth Annual ACM-SIAM Symposium on Discrete Algorithms}, pages 1180--1199. SIAM, 2020.

\bibitem{ohlmann2023canonical}
Pierre Ohlmann, Michal Pilipczuk, Wojciech Przybyszewski, and Szymon Torunczyk.
\newblock Canonical decompositions in monadically stable and bounded shrubdepth graph classes.
\newblock In {\em 50th International Colloquium on Automata, Languages, and Programming, {ICALP} 2023}, volume 261 of {\em LIPIcs}, pages 135:1--135:17. Schloss Dagstuhl - Leibniz-Zentrum f{\"{u}}r Informatik, 2023.

\bibitem{oum2006approximating}
Sang-il Oum and Paul Seymour.
\newblock Approximating clique-width and branch-width.
\newblock {\em Journal of Combinatorial Theory, Series B}, 96(4):514--528, 2006.

\bibitem{PilipczukSSTV22}
Michal Pilipczuk, Nicole Schirrmacher, Sebastian Siebertz, Szymon Torunczyk, and Alexandre Vigny.
\newblock Algorithms and data structures for first-order logic with connectivity under vertex failures.
\newblock In {\em 49th International Colloquium on Automata, Languages, and Programming, {ICALP} 2022}, volume 229 of {\em LIPIcs}, pages 102:1--102:18. Schloss Dagstuhl - Leibniz-Zentrum f{\"{u}}r Informatik, 2022.

\bibitem{robertson1983graph}
Neil Robertson and Paul~D Seymour.
\newblock Graph minors {I} -- {XXIII}.
\newblock 1983 - 2010.

\bibitem{robertson1986graph}
Neil Robertson and Paul~D Seymour.
\newblock Graph minors. {V}. excluding a planar graph.
\newblock {\em Journal of Combinatorial Theory, Series B}, 41(1):92--114, 1986.

\bibitem{robertson1995graph}
Neil Robertson and Paul~D Seymour.
\newblock Graph minors. {XIII}. the disjoint paths problem.
\newblock {\em Journal of combinatorial theory, Series B}, 63(1):65--110, 1995.

\bibitem{sau2024more}
Ignasi Sau, Giannos Stamoulis, and Dimitrios~M Thilikos.
\newblock A more accurate view of the flat wall theorem.
\newblock {\em Journal of Graph Theory}, 107(2):263--297, 2024.

\bibitem{sau2024parameterizing}
Ignasi Sau, Giannos Stamoulis, and Dimitrios~M Thilikos.
\newblock Parameterizing the quantification of cmso: model checking on minor-closed graph classes.
\newblock {\em arXiv preprint arXiv:2406.18465}, 2024.

\bibitem{schirrmacher2024model}
Nicole Schirrmacher, Sebastian Siebertz, Giannos Stamoulis, Dimitrios~M Thilikos, and Alexandre Vigny.
\newblock Model checking disjoint-paths logic on topological-minor-free graph classes.
\newblock In {\em Proceedings of the 39th Annual ACM/IEEE Symposium on Logic in Computer Science}, pages 1--12, 2024.

\bibitem{schirrmacher2023first}
Nicole Schirrmacher, Sebastian Siebertz, and Alexandre Vigny.
\newblock First-order logic with connectivity operators.
\newblock {\em ACM Transactions on Computational Logic}, 24(4):1--23, 2023.

\bibitem{schweikardt2005arithmetic}
Nicole Schweikardt.
\newblock Arithmetic, first-order logic, and counting quantifiers.
\newblock {\em ACM Transactions on Computational Logic (TOCL)}, 6(3):634--671, 2005.

\bibitem{seese1991structure}
Detlef Seese.
\newblock The structure of the models of decidable monadic theories of graphs.
\newblock {\em Annals of pure and applied logic}, 53(2):169--195, 1991.

\bibitem{seese1996linear}
Detlef Seese.
\newblock Linear time computable problems and first-order descriptions.
\newblock {\em Mathematical Structures in Computer Science}, 6(6):505--526, 1996.

\bibitem{shelah1990classification}
Saharon Shelah.
\newblock {\em Classification theory: and the number of non-isomorphic models}.
\newblock Elsevier, 1990.

\bibitem{stockmeyer1976polynomial}
Larry~J Stockmeyer.
\newblock The polynomial-time hierarchy.
\newblock {\em Theoretical Computer Science}, 3(1):1--22, 1976.

\bibitem{stockmeyer1974complexity}
Larry~Joseph Stockmeyer.
\newblock {\em The complexity of decision problems in automata theory and logic.}
\newblock PhD thesis, Massachusetts Institute of Technology, 1974.

\bibitem{thatcher1968generalized}
James~W. Thatcher and Jesse~B. Wright.
\newblock Generalized finite automata theory with an application to a decision problem of second-order logic.
\newblock {\em Mathematical systems theory}, 2(1):57--81, 1968.

\bibitem{thilikos2023excluding}
Dimitrios~M Thilikos and Sebastian Wiederrecht.
\newblock Excluding surfaces as minors in graphs.
\newblock {\em arXiv preprint arXiv:2306.01724}, 2023.

\bibitem{torunczyk2020aggregate}
Szymon Toru{\'n}czyk.
\newblock Aggregate queries on sparse databases.
\newblock In {\em Proceedings of the 39th ACM SIGMOD-SIGACT-SIGAI Symposium on Principles of Database Systems}, pages 427--443, 2020.

\bibitem{torunczyk2023flip}
Szymon Toru{\'n}czyk.
\newblock Flip-width: Cops and robber on dense graphs.
\newblock In {\em 2023 IEEE 64th Annual Symposium on Foundations of Computer Science (FOCS)}, pages 663--700. IEEE, 2023.

\bibitem{vardi1982complexity}
Moshe~Y Vardi.
\newblock The complexity of relational query languages.
\newblock In {\em Proceedings of the fourteenth annual ACM symposium on Theory of computing}, pages 137--146, 1982.

\bibitem{warwick16}
{Open problems from the workshop on algorithms, logic and structure, University of Warwick} Workshop.
\newblock 2016.

\bibitem{zhu2009colouring}
Xuding Zhu.
\newblock Colouring graphs with bounded generalized colouring number.
\newblock {\em Discrete Mathematics}, 309(18):5562--5568, 2009.

\end{thebibliography}

\end{document}